%% file: magnetic_fields_orientation.tex
\documentclass[useAMS,usenatbib]{mn2e}

\usepackage{graphicx}
\usepackage{amsmath}
\usepackage{amssymb}

\newcommand{\HII}{H\,{\sc ii} }
\newcommand{\FLASH}{{\sc flash} }
\newcommand{\disperse}{{\sc DisPerSE} }
\newcommand{\alfven}{Alfv\'{e}n }
\newcommand{\yt}{{\tt yt} }
\newcommand{\PARAMESH}{{\sc paramesh} }
\newcommand{\Msun}{M_\odot}
\newcommand{\Lsun}{L_\odot}

\renewcommand{\vec}[1]{\boldsymbol{#1}}

\title[Filamentary flow and magnetic geometry]{Filamentary flow and magnetic geometry in evolving cluster-forming molecular cloud clumps}
\author[M.\ Klassen, R.E.\ Pudritz, \& H.\ Kirk]{Mikhail Klassen$^{1}$\thanks{E-mail: klassm@mcmaster.ca}, Ralph E.\ Pudritz$^{1,2,3,4}$, Helen Kirk$^{5}$\\
$^{1}$Department of Physics and Astronomy, McMaster University, 1280 Main St.~W, Hamilton, ON L8S 4M1, Canada\\
$^{2}$Origins Institute, McMaster University, 1280 Main St.~W, Hamilton, ON L8S 4M1, Canada\\
$^{3}$Institut f\"{u}r Theoretische Astrophysik, Albert-Ueberle-Str.~2, 69120 Heidelberg, Germany\\
$^{4}$Max Planck Institut f\"{u}r Astronomie, K\"{o}nigstuhl 17, 69117 Heidelberg, Germany\\
$^{5}$National Research Council of Canada, Herzberg Astronomy and Astrophysics, 5071 West Saanich Road, Victoria, BC V9E 2E7, Canada\\
}

\begin{document}
\bibliographystyle{mn2e}

\date{4 November 2016}

\pagerange{\pageref{firstpage}--\pageref{lastpage}} \pubyear{2016}

\maketitle

\label{firstpage}

\begin{abstract}
We present an analysis of the relationship between the orientation of magnetic fields and filaments that form in 3D magnetohydrodynamic simulations of cluster-forming, turbulent molecular cloud clumps. We examine simulated cloud clumps with size scales of $L \sim 2$--$4$ pc and densities of $n \sim 400$--$1000$ cm$^{-3}$ with Alfv\'{e}n Mach numbers near unity. We simulated two cloud clumps of different masses, one in virial equilibrium, the other strongly gravitationally bound, but with the same initial turbulent velocity field and similar mass-to-flux ratio. We apply various techniques to analyze the filamentary and magnetic structure of the resulting cloud, including the \disperse filament-finding algorithm in 3D. The largest structure that forms is a 1--2 parsec-long filament, with smaller connecting sub-filaments. We find that our simulated clouds, wherein magnetic forces and turbulence are comparable, coherent orientation of the magnetic field depends on the virial parameter. Subvirial clumps undergo strong gravitational collapse and magnetic field lines are dragged with the accretion flow. We see evidence of filament-aligned flow and accretion flow onto the filament in the subvirial cloud. Magnetic fields oriented more parallel in the subvirial cloud and more perpendicular in the denser, marginally bound cloud. Radiative feedback from a 16 $\Msun$ star forming in a cluster in one of our simulations ultimately results in the destruction of the main filament, the formation of an \HII region, and the sweeping up of magnetic fields within an expanding shell at the edges of the \HII region.
\end{abstract}

\begin{keywords}
hydrodynamics -- radiative transfer -- stars: formation -- ISM: kinematics and dynamics -- ISM: magnetic fields -- ISM: structure.
\end{keywords}

\section{Introduction}\label{sec:intro}

Observations of nearby molecular clouds obtained with the \textit{Herschel Space Observatory} \citep{Pilbratt+2010} have shown them to be highly filamentary, with filaments being the clear sites for star formation \citep{Andre+2010,Henning+2010,Menshchikov+2010,Hill+2011,Polychroni+2013,Andre+2014} and the intersections of filaments the sites of clustered and massive star formation \citep{Schneider+2012,Peretto+2013}. \citet{Myers2009} presents a theoretical model based on nearby star-forming complexes of star clusters forming within ``hubs'', parsec-length filaments radiating from them like spokes. Star formation on a galactic scale is very inefficient, and so efforts to understand it have focussed on the physics of molecular clouds, which are now being imaged with unprecedented resolution. Older observations see gradients along the long axis of a filament \citep[e.g.][]{Bally+1987,Schneider+2010}, while a few new observations, such as from the CARMA Large Area Star Formation Survey (CLASSy) \citep{Storm+2014}, are high enough resolution to see velocity gradients across filaments \citep{Fernandez-Lopez+2014}.

\citet{Beuther+2015} recently imaged a massive filamentary infrared dark cloud (IRDC 18223), observed in 3.2mm continuum and in molecular line data. This massive filament has a line mass of about 1000 $\Msun$/pc. Along its length, 12 massive cores have formed with approximately even spacing. This extremely high line mass and fragmentation pattern requires additional support, either in the form of supersonic turbulence or magnetic fields.
N$_2$H$^{+}$ spectral line observations, tracing the motions of the dense filamentary gas, show significant gradients in the centroid velocity, suggesting a kinematic origin for this filament, although the authors were unable to differentiate between the potential roles played by large-scale gravitational collapse, rotation, converging magnetised gas flows, or whether the filament formed out of previously existing velocity-coherent sub-filaments.

The relevance of magnetic fields to the star formation process is already well established \citep{McKeeOstriker2007,Crutcher2012,Li+2014}, but direct measurement of magnetic fields remains difficult. Measurements of light polarization from dust grains remains the best way of mapping the plane-of-sky magnetic field orientation. Non-spherical dust grains to orient their long axis perpendicular to the ambient magnetic field \citep{Lazarian2007,HoangLazarian2008}. Starlight appears polarized parallel with the magnetic field due to absorption effects by dust \citep{Davis+Greenstein1951,Hildebrand1988}. Meanwhile, thermal emission by these dust grains produce light at far-infrared and sub-millimetre wavelengths polarized perpendicular to the magnetic field \citep{Hildebrand+1984, Novak+1997, Vaillancourt2007, Alves+2014}. This allows for the measurement of magnetic field orientations.

One of the best-studied regions of star formation is the Taurus molecular cloud. Molecular line observations of $^{12}$CO show striations aligned with the local magnetic field \citep{Heyer+2008} as traced by background starlight polarization. This observation is also confirmed by \citet{Palmeirim+2013}, who suggest that material may be accretion along these striations and onto the larger B211 filament. \citet{PlanckXXXIII} also suggest that magnetic fields affect filament formation, based on observations of polarized dust emission in the Taurus and Musca clouds.

The orientation of magnetic fields relative to the filaments in star-forming regions is of dynamical importance. One of the proposed mechanisms for the interaction between magnetic fields and filaments is that magnetic fields could channel material along the field line orientation, allowing filaments to form by gravitational contraction, as suggested by MHD simulations \citep{Nakamura+Li2008}. Low density filaments or striations should be oriented parallel to magnetic fields, channeling material onto the larger filaments \citep[cf. numerical simulation by][]{Vestuto+2003,Li+2008}.

\citet{Li+2013} examined the orientation of the filamentary giant molecular clouds of the Gould Belt ($N_H \approx 2 \times 10^{21}$--$2 \times 10^{22}$ cm$^{-2}$) relative to the magnetic fields of the intercloud medium (ICM) and found a bimodal distribution. Most clouds are oriented either perpendicular or parallel to ICM B-fields, with offsets typically less than 20 degrees. This strongly suggests the dynamical importance of magnetic fields in the formation of filaments. The physical scales observed in \citet{Li+2013} were generally a few parsecs to a few tens of parsecs, up to an order of magnitude larger than our clump-scale simulations, but of comparable density ($N_H \approx 5 \times 10^{21}$ cm$^{-2}$). The orientation of the large-scale magnetic field relative to the large-scale structure of the cloud was studied, showing a bimodal distribution with peaks near parallel and perpendicular relative orientation.

The recent publication of the Planck polarization data \citep{PlanckXXXV} shows that magnetic fields have a strong tendency to be parallel to diffuse filamentary clouds with column densities below $N_H \approx 10^{21.7}$ cm$^{-2}$, and perpendicular to dense filaments of higher column density. The authors observe a transition in relative orientation with increasing $N_H$.

Given that the magnetic field energy dominates the gravity and thermal energy in the diffuse ISM, these results clearly point to the fact in diffuse gas, magnetic fields directs infall along them, resulting in the creation of dense, self-gravitating filaments oriented perpendicular to the field on impact \citep{Beck2016}. For field-aligned flows of the more diffuse ISM arising on galactic scales, Parker instabilities can readily create the dense molecular filaments, as material flows back towards the galactic plane \citep{GomezdeCastroPudritz1992}.

Magnetic fields may also play a role in the stability of filaments. Virial analysis by \citet{FiegePudritz2000} showed that, depending on orientation, magnetic fields may work to stabilize filaments against gravitational collapse or have the opposite effect. Toroidal fields assist gravity in squeezing filaments, while the poloidal fields threading filaments offer a magnetic pressure support against gravity. While submillimetre observations of filaments have detected helical fields in some cases \citep{MatthewsWilson2000}, turbulent MHD simulations have not yet reported such structures.

Our research has focused on comparing numerical simulations of magnetised molecular cloud clumps to observations, with the aim of better understanding the co-evolution of the magnetic fields, filamentary structure, and star formation. We have examined scales on the order of a few parsec, smaller than the Gould Belt clouds examined in \citet{Li+2013}, but on the same physical scales as the Serpens South cloud \citep[e.g.][]{Sugitani+2011,Kirk+2013} and the Taurus B211 filament \citep{Palmeirim+2013}.

In \citet*{Kirk+2015} \citep[henceforth simply][]{Kirk+2015}, we examined the structure of magnetised and unmagnetised filaments via numerical simulations, showing that simulated filaments have properties consistent with observed filaments, specifically the radial column density profiles, with similar shapes and extents to those observed. Magnetic fields can offer pressure support to filaments, and were observed to be somewhat ``puffier'' than their unmagnetised counterparts, with broader radial profiles and lower central densities \citep[for details see][]{Kirk+2015}. Turbulence was also observed to play a critical role in supporting the filament against collapse, consistent with the observations by \citet{Beuther+2015} of the massive, turbulent filament. This paper is the follow-on and extension of the investigations into simulated filaments begun in \citet{Kirk+2015}.

Magnetohydrodynamic (MHD) simulations provide an experimental laboratory for studying the evolution of the turbulent ISM within a magnetic field. \citet{Soler+2013} examined the relative orientations of density gradients and magnetic fields in 3D MHD simulations with decaying turbulence. Isodensity contours serve to trace filaments in this technique. Here too, filaments were seen to lie parallel to magnetic field lines at low densities, but switch to perpendicular in high density regions. This effect was more pronounced in simulations with high magnetisation.

Characterizing the relative energies of turbulence and magnetic fields is the turbulent Alfv\'{e}n Mach number,
\begin{equation}\label{eqn:alfven_mach_number}
\mathcal{M}_A = \left(\beta/2\right)^{1/2} \mathcal{M} = \frac{\sigma}{v_A},
\end{equation}
where $\beta = 8\pi\rho \sigma^2 / \langle B^2 \rangle = 2 \sigma^2 / v_A^2$ is the plasma beta, which describes the ratio of thermal pressure to magnetic pressure. $\sigma$ and $v_A = B / \sqrt{4 \pi \rho}$ are the 1D velocity dispersion and the Alfv\'{e}n speed, respectively. The latter is the characteristic speed of a transverse magnetohydrodynamic wave. Because we are dealing with supersonic turbulence, it makes more sense to use the velocity dispersion, rather than the sound speed to characterize the gas motion. $\mathcal{M} = v / c_s$ is the thermal Mach number of the turbulence, where $v$ is the gas speed. If the turbulence is sub-Alfv\'{e}nic ($\mathcal{M}_A < 1$), turbulent pressure can elongate filaments in the direction of the magnetic field. If turbulent pressure is lacking, gravity can draw material along B-fields to form a filament with a perpendicular orientation (although perpendicular orientations can be caused by other physical processes, such as colliding flows). If the turbulence is super-Alfv\'{e}nic ($\mathcal{M}_A > 1$), then magnetic fields are not dynamically important and turbulence can compress gas in any direction to form filaments regardless of the large-scale orientation of intercloud magnetic fields. Molecular clouds are observed to possess Alfv\'{e}n Mach numbers of order unity \citep{Crutcher1999} and this is where we situate our own simulations.

\citet{Falceta-Goncalves+2008} performed MHD simulations with varying degrees of turbulence, although they do not include gravity. They examined sub-Alfv\'{e}nic ($\mathcal{M}_A = 0.7$) and super-Alfv\'{e}nic ($\mathcal{M}_A = 2.0$) cases. Field lines were ordered in the sub-Alfv\'{e}nic case, but observed to be random in the super-Alfv\'{e}nic case. They did not examine a transitional case near $\mathcal{M}_A \sim 1$. \citet{HeyerBrunt2012} looked at Alfv\'{e}nic turbulence in Taurus, and found a transition from sub-Alfv\'{e}nic turbulence in the envelope of the molecular cloud to super-Alfv\'{e}nic within the denser regions of the cloud. Additionally, the envelope showed a velocity anisotropy aligned with the local magnetic field. Our simulations of trans-Alfv\'{e}nic clouds are therefore well-situated and many molecular clouds are observed to have Alfv\'{e}n Mach number close to unity \citep{Crutcher1999}.

We perform numerical simulations of rotating, magnetised, turbulent molecular cloud clumps and the study the evolution of the resulting network of filaments. The aim is to better understand the competing influences of turbulence, magnetic fields, gravity and stellar feedback. Our two simulations are initialized as spherically symmetric in density, with magnetic fields oriented parallel to the $z$-axis, and then allowed to evolve. They differ mainly in terms of gravitational boundedness, parametrized by the virial parameter: one cloud clump is marginally bound, the other is highly bound and therefore collapses rapidly due to gravity. We explore the relationship between gravitational boundedness and the relative orientation of filaments and the magnetic field. Finally, in one simulation, we include massive star formation, which injects energy back into the densest parts of the main filament via ionizing radiation. We examine how this affects the magnetic field and accretion flows along the primary filament.

In the following sections we first describe our numerical methods (Section \ref{sec:numerical_methods}), lay out our results for filamentary and magnetic structure during early cloud evolution (Section \ref{sec:results}), analyze the effects of radiative feedback during later cloud evolution (Section \ref{sec:radiative_feedback}), and then discuss our results (Section \ref{sec:discussion}).

\section{Numerical Methods}\label{sec:numerical_methods}

\input{table_sim_params.tex}

We explore the relationship between the structure of molecular gas inside cloud clumps and the presence of magnetic fields through numerical simulation. MHD simulations, initialized with a turbulent velocity field, allow us to explore the three-dimensional structure of turbulence and filaments. Projections can then be made of this three-dimensional data to produce column density maps for closer comparison to astronomical observations.

We consider two approaches to studying the relative orientation of the magnetic field and filaments. The first, inspired by \citet{Li+2013}, considers the ``mean'' molecular cloud orientation in column density projection and the magnetic field orientation relative to this axis. The second approach we take considers the simulated volume in 3D, and maps out the 3D filamentary structure using the \disperse algorithm\footnote{http://www2.iap.fr/users/sousbie/} \citep{Sousbie1,Sousbie2}, and then measures the relative angle formed by the filament and the local magnetic field at locations along each filament.

Through the application of structure-mapping algorithms, we can extract the (2D or 3D) filamentary networks evident in the simulation data (from column or volume density). Once the filamentary structure is mapped, we can compare local magnetic field orientation to study how orientation might be related filament characteristics.

The analysis of \textit{Herschel} results has seen the widespread application of image analysis algorithms for filament detection. These include the \textit{getfilaments} algorithm by \citet{getfilaments} and \disperse \citep{Sousbie1,Sousbie2}. 

We opted to use \disperse, which maps out the topological features in an image or datacube, such as peaks, voids, or filaments. It has the advantage of being applicable also to 3D data cubes built from our simulation data. Previously, \disperse had seen wide use for analysis of {\it Herschel} observations \citep[e.g.][]{Arzoumanian+2011,Peretto+2012,Hennemann+2012,Schneider+2012}, but we used it in a new way---on 3D star formation simulation data for examining the 3D structure of molecular gas filaments. This is in contrast to analyzing filaments solely in projection and is one of the first times this has been done in 3D for studies of filaments inside molecular clouds \citep[see][for another example]{Smith+2014}.

\subsection{Numerical simulations}\label{sec:numerical_simulations}

We perform numerical magnetohydrodynamic simulations using the \FLASH AMR code \citep{Fryxell2000} in its version 2.5. It makes use of the \PARAMESH library to solve the fluid equations on an adaptive Eulerian grid \citep{Olson+1999,MacNeice+2000}. The code has been expanded to include Lagrangian sink particles \citep{Banerjee2009, Federrath2010}, radiative heating and ionization feedback \citep{Rijkhorst2006, Peters2010a}, and self-consistent protostellar evolution \citep{Klassen+2012a}.

\subsection{Initial conditions}\label{sec:initial_conditions}

We performed numerical simulations, which we label {\tt MHD500} and {\tt MHD1200}, of two molecular clouds at different scales, one ({\tt MHD500}) more compact and close to virial equilibrium ($\alpha_{\mathrm{vir}} = 0.95$), with approximately $500 \Msun$ of material in a volume of side length 2.0 pc, and the second ({\tt MHD1200}) containing about $1200 \Msun$ of material in a volume of side length 3.89 pc and considerably more subvirial ($\alpha_{\mathrm{vir}} = 0.56$). These molecular clouds have column densities similar to those observed for Gould Belt clouds. The column density of the {\tt MHD500} cloud has an average value of $N_H = 7.33 \times 10^{21}$ cm$^{-2}$ and a peak value of $N_H = 3.4 \times 10^{23}$ cm$^{-2}$ at the start of the simulation. The {\tt MHD1200} cloud has an average column density of $N_H = 4.66 \times 10^{21}$ cm$^{-2}$ and a peak value of $N_H = 2.13 \times 10^{23}$ cm$^{-2}$ at the start of the simulation. Comparing to Figure 7 from \citet{Li+2013}, we see that the average value of the column densities and the magnetic fields correspond to Gould Belt clouds, while the peak column densities correspond to cloud cores. For comparison, the column density estimates for Gould Belt clouds in \citet{PlanckXXXV} were $N_H \approx 1$--$10 \times 10^{21}$ cm$^{-2}$ for average column densities, with peak values of $N_H \approx 20$--$100 \times 10^{21}$ cm$^{-2}$. Additionally, \textit{Herschel} observations towards Aquila and Polaris, showed column densities within star-forming filaments of around 1--2 $\times 10^{22}$ cm$^{-2}$ in Aquila, with non-star-forming filaments in both clouds showing column densities up to a few $10^{21}$ cm$^{-2}$ \citep{Andre+2010}.

The initial conditions are very similar to those we performed for \citet{Kirk+2015}, where the reader may find further details. The {\tt MHD500} simulation is the same in both papers. Previously, we examined filaments properties, comparing purely hydrodynamic simulations with MHD simulations. In this paper, we focus on magnetised molecular cloud clumps, and include a simulation ({\tt MHD1200}) that has star formation and photoionizing feedback in order to study the effect this form of radiative feedback has on filamentary structure and the magnetic field orientation. In \citet{Kirk+2015} we did not include radiative feedback as part of our study.

Our initial conditions are listed in Table \ref{table:simulation_parameters}. These types of simulations are computationally expensive to perform, and in this paper we limit ourselves to examining two simulated molecular cloud clumps with different average gas densities, but comparable magnetisation. The boundary conditions of our volume were open (outflow condition), but this does not influence the evolution in any significant way, as the clouds collapse gravitationally. Our grid resolution for the {\tt MHD500} simulation was about 50 AU, and for our {\tt MHD1200} it was about 390 AU. This was on account of the larger box size in the latter case. This resolution is sufficient to resolve filaments, which in \citet{Kirk+2015} were found to be between 0.06 pc ($\approx 12000$ AU) and 0.26 pc ($\approx 54000$ AU).

Molecular clouds are observed to have non-thermal linewidths attributed to supersonic turbulence \citep{Larson1981,Larson2003}. We initialize our simulations with a turbulent velocity field that is a mixture of compressive and solenoidal modes with a Burgers spectrum, $E_k \propto k^{-2}$ \citep{Federrath+2008,Girichidis2011}. Kolmogorov turbulence ($E_k \propto k^{-5/3}$) would be expected for incompressible fluids. The largest modes in our simulations have characteristic size scales on the order of box width (3.89 pc and 2 pc, for our {\tt MHD1200} and {\tt MHD500} simulations, respectively). See also \citet{Larson1981,Boldyrev2002,HeyerBrunt2004}. The randomly-oriented velocities sampled from the turbulent velocity distribution, represent only an initial condition and are allowed to decay. They are in no way correlated to the initial magnetic field orientation, which is uniform and parallel to the $z$-axis. We use the same turbulent initial velocity field for both simulations so that the structures arising from the turbulence will be similar. The root-mean-square (RMS) velocity and Mach number of each simulation is the same, and we initialized the simulations with root-mean-square velocities equal to 6 times the isothermal sound speed, but the mass-weighted average Mach number between our two simulations differs (see Table \ref{table:simulation_parameters}). 

Observations of dense molecular cloud cores forming high-mass stars show column density profiles consistent with power laws \citep{Pirogov2009}. Hence, we initialize our simulations with density profiles of the form $\rho(r) \propto r^{-3/2}$. We simulated two different initial conditions, looking at a high mass case (1200 $\Msun$) and a low mass case ($500 \Msun$). The low-mass simulation was run without any radiative feedback, whereas in our high-mass simulation we allowed stars to influence their environments via ionizing feedback.

Observations of molecular cloud cores with sizes in the range of 0.3--2.1 pc and masses up to several thousand $\Msun$ show velocity gradients consistent with a ratio of rotational to gravitational energy, $\beta_{\textrm{rot}} \lesssim 7\%$ \citep{Pirogov+2003}. Numerical simulations of molecular clouds are sometimes initialized in rigid body rotation a low $\beta_{\textrm{rot}}$ \citep[e.g.][]{Peters+2010}.

Our molecular cloud clumps are initialized in slow rigid body rotation about the $z$-axis. The rotation rate is set to $\Omega = 1.114 \times 10^{-14}$ s$^{-1}$ in both cases, which corresponds to a ratio of rotational kinetic to gravitational binding energy, $\beta_{\textrm{rot}}$ of 1\% in the {\tt MHD500} simulation and 3.2\% in the {\tt MHD1200} simulation.

The magnetic field is initially uniform and oriented parallel to the $z$-axis. We quantify magnetic field strengths via a mass-to-flux ratio, normalized against a critical mass-to-flux ratio:

\begin{equation}\label{eqn:mass_to_flux}
\lambda = \frac{M/\Phi}{\left(M/\Phi\right)_{\textrm{crit}}}
\end{equation}

\citet{Crutcher2010} finds molecular clouds to typically have magnetic field strengths such that $\lambda \sim$ 2--3, so we initialize our simulations with the values in this range. The mass-to-flux ratio quantifies the dynamical importance of the magnetic field relative to gravity. The critical mass-to-flux ratio ($\lambda_{\textrm{crit}} \approx 0.13/\sqrt{G}$) is the value needed for gravitational energy to be balanced by the magnetic energy \citep{MouschoviasSpitzer1976}. The upper range for massive star forming regions is $\lambda \lesssim 5$ \citep{Falgarone+2008,Girart+2009,Beuther+2010}.

Using equation \ref{eqn:alfven_mach_number}, we calculate the RMS \alfven Mach number of each simulation, an important measure into whether the magnetic fields will dominate the turbulence. The {\tt MHD500} cloud has an RMS \alfven Mach number of 0.92, while the {\tt MHD1200} cloud has an RMS \alfven Mach number of 0.99, i.e.~both clouds have \alfven Mach number very close to unity and are essentially trans-Alfv\'{e}nic, meaning they are in the regime where turbulence threatens to destroy any orderly magnetic field structure the cloud may have inherited from the ICM. This regime is of interest because most clouds have \alfven Mach numbers close to unity \citet{Crutcher1999}.

Finally, we compare the kinetic energy to the gravitational energy by calculating the virial parameter \citep{BertoldiMcKee1992} for each of our clouds. The virial parameter,
\begin{equation}\label{eqn:virial_parameter}
\alpha_{\textrm{vir}} = \frac{2 \mathcal{T}}{|\mathcal{W}|} \approx \frac{5 \sigma^2 R}{G M},
\end{equation}
where $R$ is the radius of the cloud, $G$ is Newton's constant, $M$ is the mass of the cloud, and $\sigma$ is the velocity dispersion, usually measured from line width observations, measures the boundedness of the clouds. Most clouds have virial parameters of $\alpha_{\mathrm{vir}} = 0.5$--$5$ \citep[see Figure 6,][]{Rosolowsky2007} and a cloud with $\alpha_{\textrm{vir}} < 1$ is expected to collapse gravitationally, while clouds with $\alpha_{\textrm{vir}} \approx 1$ are marginally stable against collapse. We find the one-dimensional velocity dispersion by taking the mass-weighted average velocity in our simulation,
\begin{equation}
\sigma = \left(\frac{\int \rho(\vec{r})|\vec{v}(\vec{r})|^2 dV}{3 \int \rho(\vec{r}) dV} \right)^{1/2},
\end{equation}
where the number $3$ in the denominator is the geometrical factor accounting for the number of dimensions. The virial parameter of the {\tt MHD500} simulation is 0.95, i.e.~marginally bound, while the {\tt MHD1200} is substantially sub-virial at $\alpha_{\textrm{vir}} = 0.56$.

We provide a summary of our simulation parameters in Table \ref{table:simulation_parameters}.

\subsection{Filament-finding}\label{sec:disperse}

Filamentary structure results from collisions between supersonic shocks \citep{MacLowKlessen2004,Schneider+2011,PudritzKevlahan2013}. Once the simulations are sufficiently evolved, filamentary structure is well developed. This always preceeds any star formation within the cloud clump. We took the evolved simulation output for our filaments analysis.

Identifying filamentary structure is challenging, and a variety of techniques have been developed for this task. One of the most straightforward approaches is based on structure-characterisation. \citet{Hennebelle2013} set characteristic density thresholds for molecular clumps. Filaments represented elongated clumps. In \citet{PlanckXXXII}, a Hessian matrix is defined for every pixel of the dust intensity map. By solving for the eigenvalues of this matrix, the local curvature is defined and filamentary structure can be extracted. The authors then construct a mask based on the intensity contrast relative to the background dust map, curvature, and the signal-to-noise of the polarization fraction. This selects the most significant ridge-like structures in the all-sky \textit{Planck} map.


Another approach is the ``histogram of relative orientations'' (HRO) developed by \citet{Soler+2013}, which is based on a computer vision algorithm called the Histogram of Oriented Gradients. Gradients in either the volume density or column density are used to indicate filaments, as filaments must lie perpendicular to the gradient vector. The relative angle between the gradient and the magnetic field orientation may then be used as a proxy for the relative orientation of magnetic fields and filaments. If the magnetic field is parallel to the density gradient, it thus lies perpendicular to the filament.

Applying this technique to magnetohydrodynamic simulations, they showed how there exists a threshold density above which the relative orientation switches from parallel to perpendicular. This threshold density was dependent on the magnetic field. The technique was also applied by the Planck collaboration for the analysis of Gould Belt clouds using the polarization of thermal dust emission observed by the {\it Planck} satellite at 353 GHz \citep{PlanckXXXV}. Gradients were measured in the column density maps and compared to the magnetic orientation inferred from polarimetry. They also found that magnetic fields went from having mostly parallel or random orientation at low density, to mostly perpendicular orientation at high density.

The Planck Collaboration is the current state-of-the-art in mapping the relative orientation of galactic magnetic fields around nearby clouds, and has added greatly to the available measurements and statistics of dust polarisation \citep{PlanckXXXII, PlanckXXXIII, PlanckXXXV}.

One disadvantage of the HRO method is that density gradients do not, by definition, imply the presence of filaments. Cores, sheets, and bubbles are coherent structures with density gradients that one would want to exclude from a study of filaments. \disperse can be sensitive to noise, but will only select filamentary structure. The use of \disperse in 3D magnetohydrodynamic calculations is also a valuable complement to extensive observational surveys.

For most of this paper, we focus on filaments and the local magnetic field structure, although Section \ref{sec:radiative_feedback} treats the subject of radiative feedback and its disruption of filamentary structure. This work also follows up on our successful use of \disperse to map structure in filaments in \citet{Kirk+2015}, with one main difference being the application of \disperse in 3D to volume density cubes instead of in 2D to column density projections. This allows us to avoid projections effects that could, for instance, make sheets appears as filaments in 2D.

In this paper we have borrowed from techniques used by \citet{Li+2013} for revealing large scale structure, and the \disperse algorithm for identifying individual filaments. \disperse has seen wide application in the analysis of {\it Herschel} results \citep[see, e.g.][]{Arzoumanian+2011,Hill+2011,Schneider+2011,Peretto+2012}, and applied to 3D hydrodynamic simulations in \citet{Smith+2014}. The analysis of 3D data is necessarily more complicated and performing ``by-eye'' assessments more complicated. There is also currently no way of ensuring that the filament skeletons extracted from the simulation grid at discrete points in the time map to the same structures in 3D. These filaments are constantly moving, shifting, and evolving. They might merge or dissipate or become disrupted by stellar feedback. 

While the filament skeletons extracted by \disperse are sensitive to the input parameters (lower persistence or noise thresholds tend to identify more striations), these structures are real topological features in the volumetric density maps. The properties of simulated filaments were compared to observed filaments in our earlier study \citep{Kirk+2015} and found to be in agreement.

\FLASH uses an adaptive mesh composed of blocks containing $8\times8\times8$ cells. The grid is refined as needed to resolve the gravitational collapse. \FLASH writes simulation plot files at specified intervals containing information about the state of the simulation and the values of grid variables. These represent global state of the simulation at a discrete point in time and are the primary output that we analyze. We refer to plot files often throughout this text.

Hierarchical grid structure is preserved in these plot files using the HDF5 file format. We used \yt \citep{ytpaper}, a general-purpose data analysis tool for computational astrophysics, to load the \FLASH output data and resample it to a uniform grid. Memory constraints meant that we mapped the density information to a $256\times256\times256$ uniform grid and wrote the output to a FITS file, a format compatible with \disperse. The remapping to a uniform grid results in the loss of some information at the highest gas densities but preserves the large-scale structure throughout the simulation volume. The remapping is necessary because \disperse does not currently support hierarchical grid structures. At a $256^3$ sampling, the grid resolution is approximately 1600 AU (0.008 pc) for our {\tt MHD500} simulation and 3100 AU (0.02 pc) for our {\tt MHD1200} simulation.

It is on these remapped FITS files that \disperse operates, first to generate the Morse-Smale complex, then to extract the filament skeleton. The Morse-Smale complex is computed by finding all the critical points (where $\nabla \rho(x,y,z) = 0$). The maxima define a set of descending manifolds (the regions where all integral lines traveling along the gradients share the same maximum), while the minima define a dual set of ascending manifolds (the integral lines all share the same minimum). The intersection of these two manifolds defines a new set that is called the Morse-Smale complex. The simulation volume is partitioned into a natural tesselation of cells. The line segments connecting maxima and passing through saddle points are a natural way to define filaments. In cosmological studies, connecting dark matter haloes in the way allows for mapping of the cosmic web \citep{Sousbie2}. In molecular clouds, we use it to trace filaments. For visualization and analysis, we can use these filament skeletons together with the original un-resampled HDF output files from FLASH to retrieve other variables of interest along their extent (e.g.~magnetic field information).

We apply this process to several plot files from each simulation. The calculation of the Morse-Smale complex is particularly computationally intensive. To handle this, we used Amazon's Elastic Compute Cloud (EC2), especially their {\tt c3.8xlarge}-type compute-optimized instances which provide 32 virtual CPUs and 60 GiBs of attached memory on demand. \disperse can then be run in parallel across these cores.

In order to avoid tracing filaments in the noise of the data, and in order to select only the most prominent filaments, \disperse measures the ``persistence'' of topological structures. Local maxima and minima form pairs of critical points. The absolute difference in value between these two is the persistence. A persistence cut removes pairs below a given threshold, but the topology of structures consisting of high-persistence points remains. By selecting the appropriate persistence threshold, we can avoid tracing filaments within the noise. ``Noise'' in our simulations would actually be small-scale density perturbations resulting from the high-wavenumber part of the turbulent power spectrum. These might appear as striations, lumps, or voids.

Another challenge when it comes to visualizing this data in 3D was that filaments span a large dynamic range in density. Because we simulated molecular cloud cores with a power-law density distribution, as are typically observed, and superimposed a supersonic velocity field on top of this, 3D plots of the filamentary structure require the selection of isosurfaces at particular densities. A volume rendering approach using raytracing requires a particular transfer function designed to highlight gas at particular densities. When the filament changes density along its length it can be difficult to plot the morphology using standard techniques.

To give an observer's picture, we take the output from these evolved simulations and project the density along each of the coordinate axes to produce column density maps from different perspectives. We then project the 3D filamentary skeletons onto these column density maps. Often there is clear agreement between the filament maps and the column density maps. Other times, the filamentary structure found in 3D is not obviously visible in 2D projection. 

To produce a plane-of-sky magnetic field map, we perform a density-weighted projection along the same coordinate axes. We project the components of the magnetic field that lie perpendicular to the axis of projection. This gives us, at every location in our map, the integrated local magnetic field orientation. By weighting the magnetic field projection by density, we favour contributions to the magnetic field orientations local to filaments in the line of sight. This is, of course, an imperfect proxy for true polarization measures, but in the case of emission, assuming dust grains of homogeneous size and alignment efficiency, light polarization would be weighted by dust density, which traces the gas density. 

To measure the orientation of the magnetic field relative to the filament, we trace the filament skeletons and interpolate $B_x$, $B_y$, and $B_z$ to find the local magnetic field orientation at the filament spine. We then measure the angle formed by the unit vectors of filament orientation and magnetic field orientation.

\section{Filament and B-Field Orientations: Early Cloud Evolution}\label{sec:results}

\begin{figure*}
\includegraphics[width=1.0\textwidth]{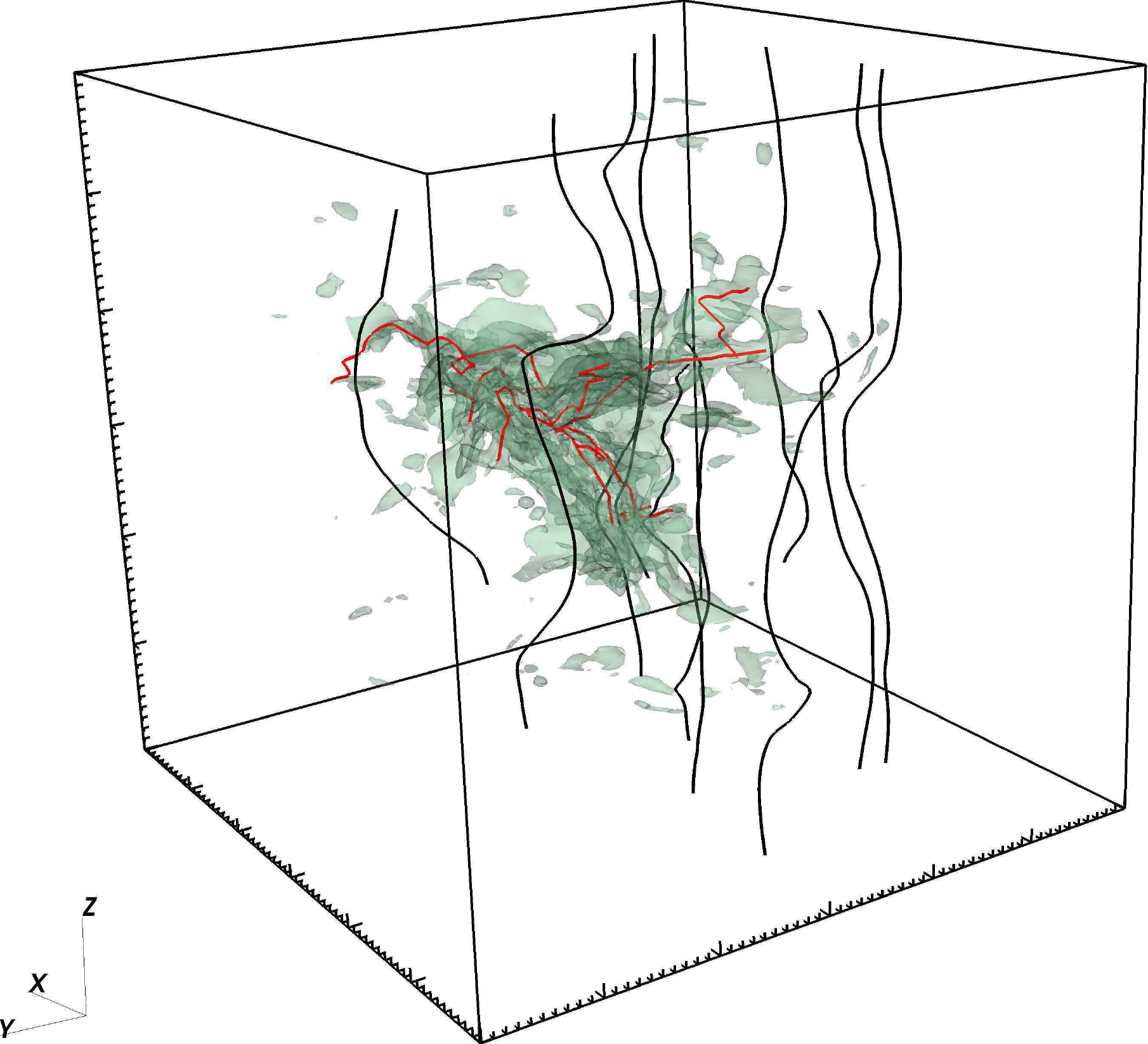}
\caption{3D plot of gas density from the {\tt MHD1200} simulation at 250,000 years of evolution. Green isosurfaces indicate gas at densities of $n = 3.1 \times 10^3$ cm$^{-3}$ ($\rho = 1.1 \times 10^{-20}$ g/cm$^3$). Red lines indicate filament skeleton selected by \disperse. Black lines are magnetic field lines at 8 randomly selected locations within the volume.}
\label{fig:m1200_whole_box}
\end{figure*}

In this section, we examine cloud and magnetic field properties during the first 250 kyr of cloud evolution before radiative feedback becomes important. 

In Figure \ref{fig:m1200_whole_box}, we take a representative plotfile from the {\tt MHD1200} simulation, after just over 250 kyr of evolution. We ran \disperse with a persistence threshold of $10^{-17}$ g/cm$^3$, which selects some of the major-trunk filaments within the volume. Gas below a density of $10^{-22}$ g/cm$^3$ is excluded from consideration. The persistence threshold was adjusted manually until only major filaments were being selected. Small-scale turbulence can cause \disperse to identify many potentially spurious short-length filaments that we did not wish to include in our analysis. A higher persistence threshold removes these from consideration.

Figure \ref{fig:m1200_whole_box} highlights in green the  $n = 3.1 \times 10^3$ cm$^{-3}$ ($\rho = 1.1 \times 10^{-20}$ g/cm$^3$) isosurface. The box enclosing the rendering depicts the entire simulation volume, with a side length of 3.89 pc.

Traced in red are the main filaments as \disperse identifies them, satisfying the selection criteria described above. They align with some of the obvious filamentary structure visible in the rendering.

To visualize what is happening with the magnetic field, we draw magnetic field lines that trace the orientation of the magnetic field from 8 randomly sampled locations. Recall that the magnetic field is initially parallel to the $z$-axis of the simulation volume. Over 250 kyr, the structure of the magnetic field has evolved in response somewhat to the slow rigid body rotation of the gas about the $z$-axis, but much more to the turbulent velocity field. The \alfven Mach number quantifies the relative energies of the turbulence and magnetic fields. Our simulated clouds have \alfven Mach numbers of approximately unity, indicating an approximate equipartition in energies. By contrast the energy in rotation is only a few percent of the gravitational binding energy, which is greater even than the kinetic energy in the {\tt MHD1200} simulation. Figure \ref{fig:m1200_whole_box} shows large deflections in the magnetic field from the initial orientation.

\begin{figure*}
\includegraphics[width=1.0\textwidth]{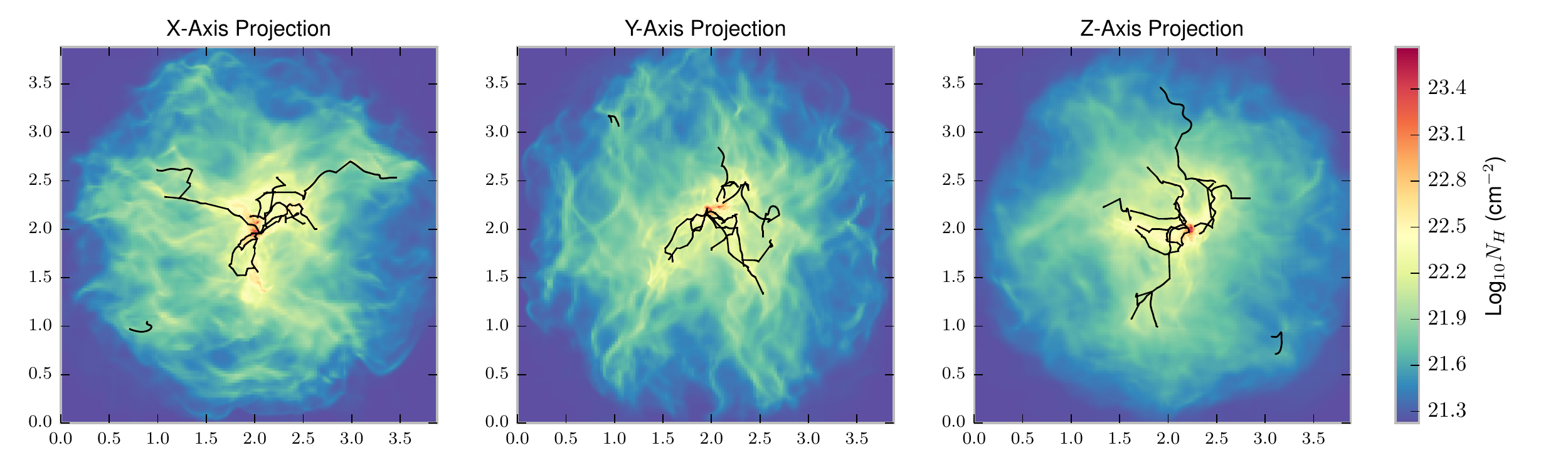}
\caption{2D column density maps along each of the coordinate axes with projections of the filament skeleton overplotted. Data is same as in Figure \ref{fig:m1200_whole_box}}
\label{fig:coldens_projections}
\end{figure*}

While Figure \ref{fig:m1200_whole_box} illustrates the structure of the magnetic field and filaments, we use projections to confirm whether structures seen in projection align with the filamentary structure that is traced in 3D. Studies of the filamentary nature of molecular clouds rely on some proxy of the column density (dust emission or integrated line intensity over some range of velocities). Density is seen in projection and hence may be hiding important structure inside the third dimension. Efforts to extract information about the 3D structure of molecular gas often make use spectral velocity data, with examples from simulations \citep[e.g.][]{Ward+2012} and observations \citep{Hacar+2011,Hacar+2013}. In the latter case it was found that filaments often have coherent velocity structures, with subsonic velocity dispersions, and marginal stability. In \citet{Kirk+2015} we also noted evidence of this kind of fine structure in numerical simulations. 

In Figure \ref{fig:coldens_projections} we take the same data as in Figure \ref{fig:m1200_whole_box}, but in order to verify that apparent structures in projection match the structures found by \disperse in 3D we also project the filament skeleton into each of the three coordinate axes and plot them side-by-side. Column densities range from about 0.01 g/cm$^2$ to about 1.0 g/cm$^2$, while the mean initial column density of the simulation was 0.02 g/cm$^2$. The black lines in Figure \ref{fig:coldens_projections} indicate the projected filaments. We see that most of the major structures are captured by this technique. This confirms that at least some of the observed major structures in the column density plots are not just the result of projection effects, but correspond to true filamentary structures in 3D. \citet{Smith+2014} also found correspondence when comparing 2D and 3D {\sc DisPerSE}-mapped filaments from simulations, except that filaments seen in column density projection did not belong to a single structure, but where made up of a network of sub-filaments reminiscent of those observed by \citet{Hacar+2013}.

\subsection{Large-scale magnetic field orientation}\label{sec:large_scale_structure}

\begin{figure*}
\includegraphics[width=1.0\textwidth]{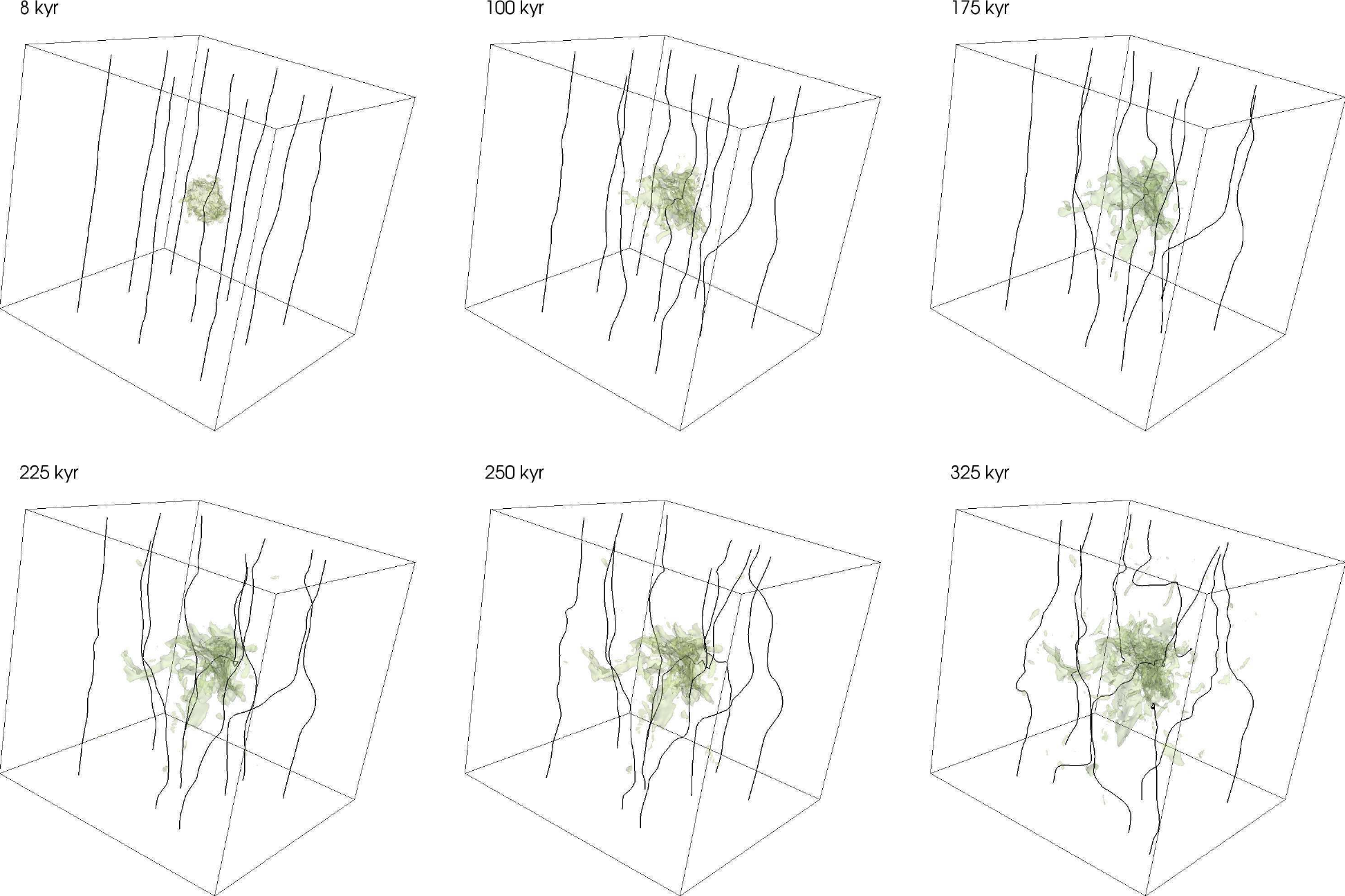}
\caption{B-field streamline evolution over the course of 325 kyr in our {\tt MHD1200} simulation. The green density contour is as in Figure \ref{fig:m1200_whole_box}. Box depicts entire simulation volume with $L =  3.89$ pc on a side. Turbulence and rotation largely account for the changes in local magnetic field orientation.}
\label{fig:m1200_streamline_evolution}
\end{figure*}

%
In Figure \ref{fig:m1200_whole_box}, in which the magnetic field lines show significant deformation, the structure of the magnetic field is the result of gas motions: the initial solid body rotation of the molecular cloud clump, the turbulent velocity field, and the slow gravitational collapse. In ideal MHD, a good assumption for the interstellar medium, fluid is assumed to conduct perfectly, and so magnetic field lines are dragged along with the fluid. 

We show the evolution of this field line dragging through the series of panels in Figure \ref{fig:m1200_streamline_evolution} in which we have plotted the magnetic field lines at 6 different snapshots in time, with the density isocontour of $\rho = 1.1\times10^{-20}$ g/cm$^3$ ($n = 3.1 \times 10^3$ cm$^{-3}$) highlighted, as in Figure \ref{fig:m1200_whole_box}. The first panel shows the state of the simulation at 8 kyr. The magnetic field lines are almost perfectly parallel with the $z$-axis, which was the initial condition. In the next panel, at 100 kyr, their state reflects some field line dragging due to the turbulent velocity field and solid body rotation. The deformation becomes more and more extreme from one panel to the next, with the last panel at 325 kyr reflecting very little of the original structure of the magnetic field.


\begin{figure*}
\includegraphics[width=1.0\textwidth]{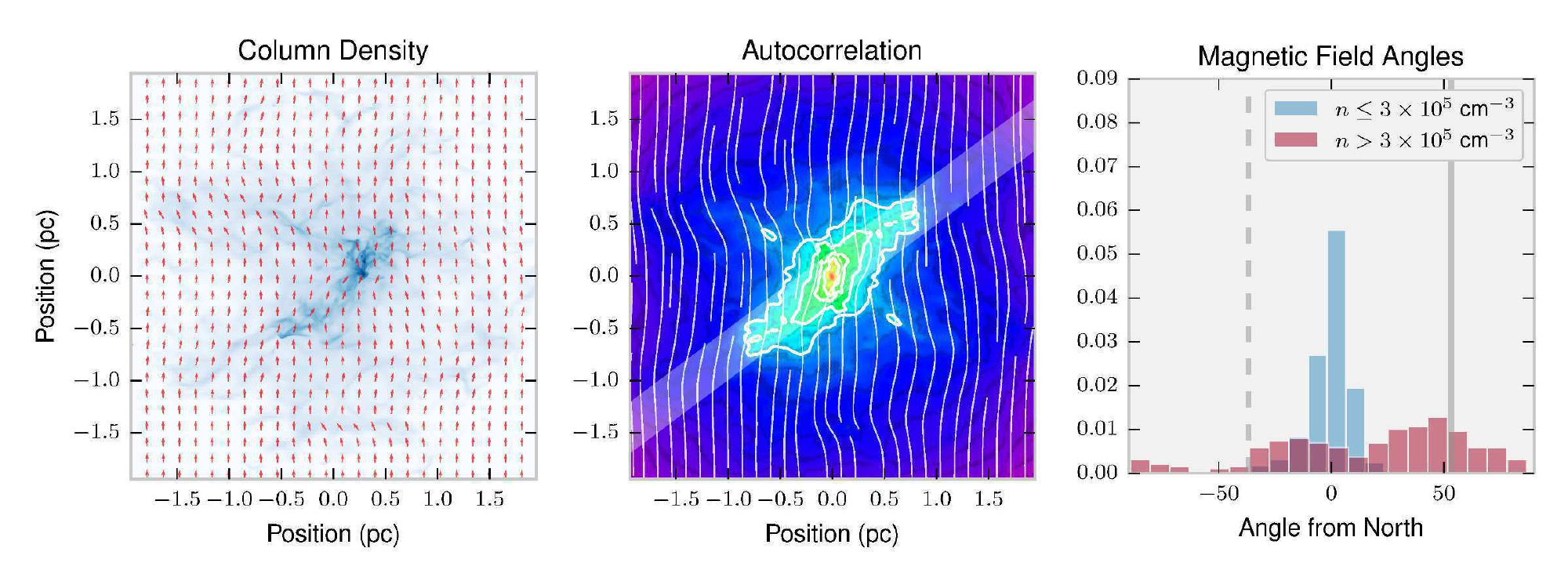}
\caption{{\it Left:} Column density maps along the $y$-axis of our {\tt MHD1200} simulation at 250 kyr of evolution with density-weighted projected magnetic field orientation overplotted in red arrows. {\it Middle:} The autocorrelation of the previous column density map, which highlight self-similar structure. Contours highlight levels from 1\% to 10\% of peak correlation values. The diagonal bar represents cloud orientation and is the best fit line through the pixels contained within the outermost contour, weighted by the base-10 logarithm of the autocorrelation values. Magnetic field lines based on the values measured for the left panel are overplotted. {\it Right:} The histogram of magnetic field orientations based on the values measured for the left panel, normalized so that the total area is 1. The orientations are measured relative to ``north''. The blue histogram shows the total distribution, while red indicates only the orientations of high-density gas. The vertical grey lines indicate the angle of the best fit line to the large-scale structure (solid right line), and the angle offset by 90$^\circ$ (dashed left line).}
\label{fig:m1200_large_scale_orientation}
\end{figure*}

\begin{figure*}
\includegraphics[width=1.0\textwidth]{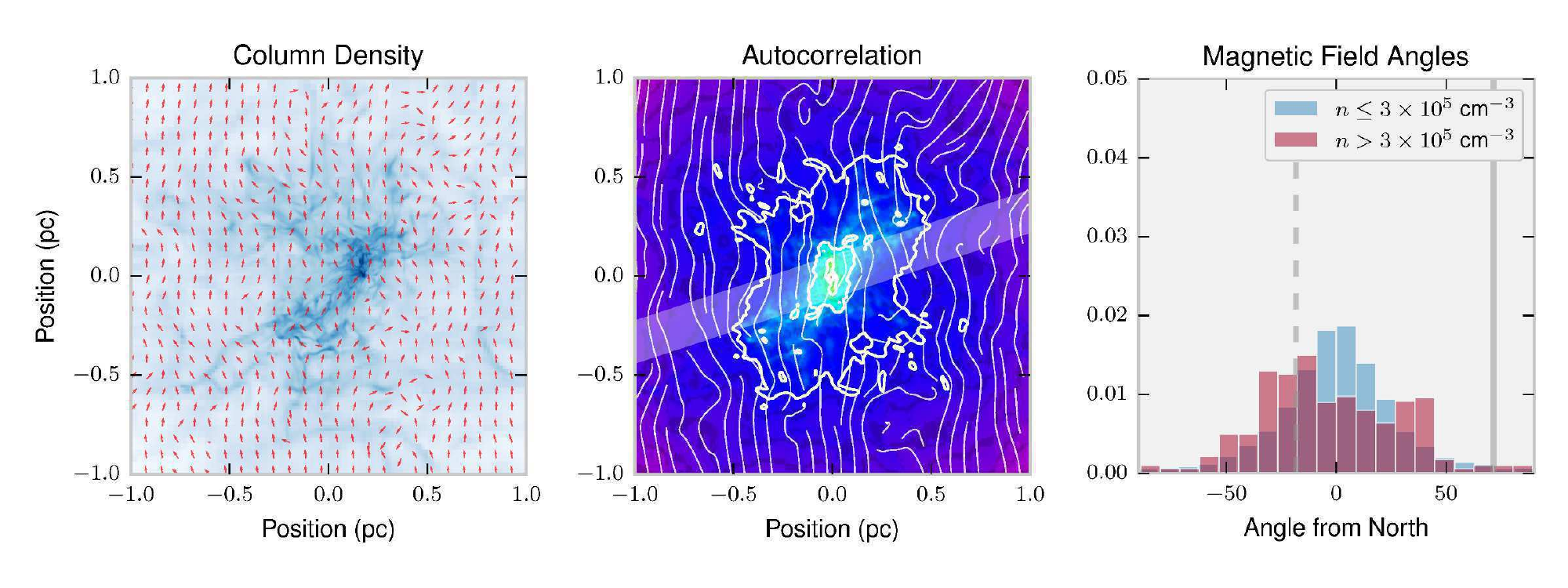}
\caption{The same as in Figure \ref{fig:m1200_large_scale_orientation}, except using {\tt MHD500} at 150 kyr of evolution. Because of the higher average gas density, we draw contours in the middle panel from 0.1\% to 10\% of peak correlation values. The best-fit line is still calculated based on the outermost contour. The data is taken at approximately the same number of freefall times as in the {\tt MHD1200} simulation.}
\label{fig:m500_large_scale_orientation}
\end{figure*}

\subsection{Magnetic fields in 2D}

In \citet{Li+2013}, the authors took observations of molecular clouds in the Gould Belt with physical sizes of a few to a few tens of pc. Many of these clouds have large-scale elongated structure. The authors were interested in the magnetic field orientation relative to orientation of the large-scale structure of the cloud. Plotting these clouds in galactic coordinates (see their Figures 1 and 2), they then took the autocorrelation of the extinction maps. The autocorrelation map is produced by taking a copy of the image and displacing it in $x$ and $y$ coordinates, and at each such displacement calculating the sum of the product of the overlapping pixels. The map of this as one image is shifted relative to its copy highlights any self-similar structure. Any elongated structures feature prominently in the autocorrelation maps. \citet{Li+2013} then contoured various levels in the autocorrelation maps and took the best-fit linear regression to the pixel positions in the contours, giving the long-axis orientation of the molecular cloud. The angle of this could then be compared to the angles of magnetic field measurements derived from polarimetry data.

Motivated by the \citet{Li+2013} approach, we perform a similar analysis on our simulation data. In Figure \ref{fig:m1200_large_scale_orientation} we show data from our {\tt MHD1200} simulation at 250 kyr of evolution, while in Figure \ref{fig:m500_large_scale_orientation} we show data from our {\tt MHD500} simulation after 150 kyr of evolution. Although taken at different times, the data are at the same number of freefall times in each simulation ($t_{\textrm{ff}} \approx 0.14$). The freefall time,
\begin{equation}
t_{\textrm{ff}} = \sqrt{\frac{3 \pi}{32 G \bar{\rho}}},
\end{equation}
is a natural measure of the gravitational timescale of the simulation. The first panel from the left shows the map of projected mean density. This is very similar to a column density projection, except that each sample in the integral is itself weighted by density. We opted for this approach because it brings dense structures into stronger relief and produces clearer autocorrelation maps. We then overplot the magnetic field vectors in red. The magnetic field values are density-weighted averages computed by integrating through the simulation volume along the same projection axis. Unlike polarimetric observations, we can measure the magnetic field everywhere.

\begin{figure*}
\includegraphics[width=0.7\textwidth]{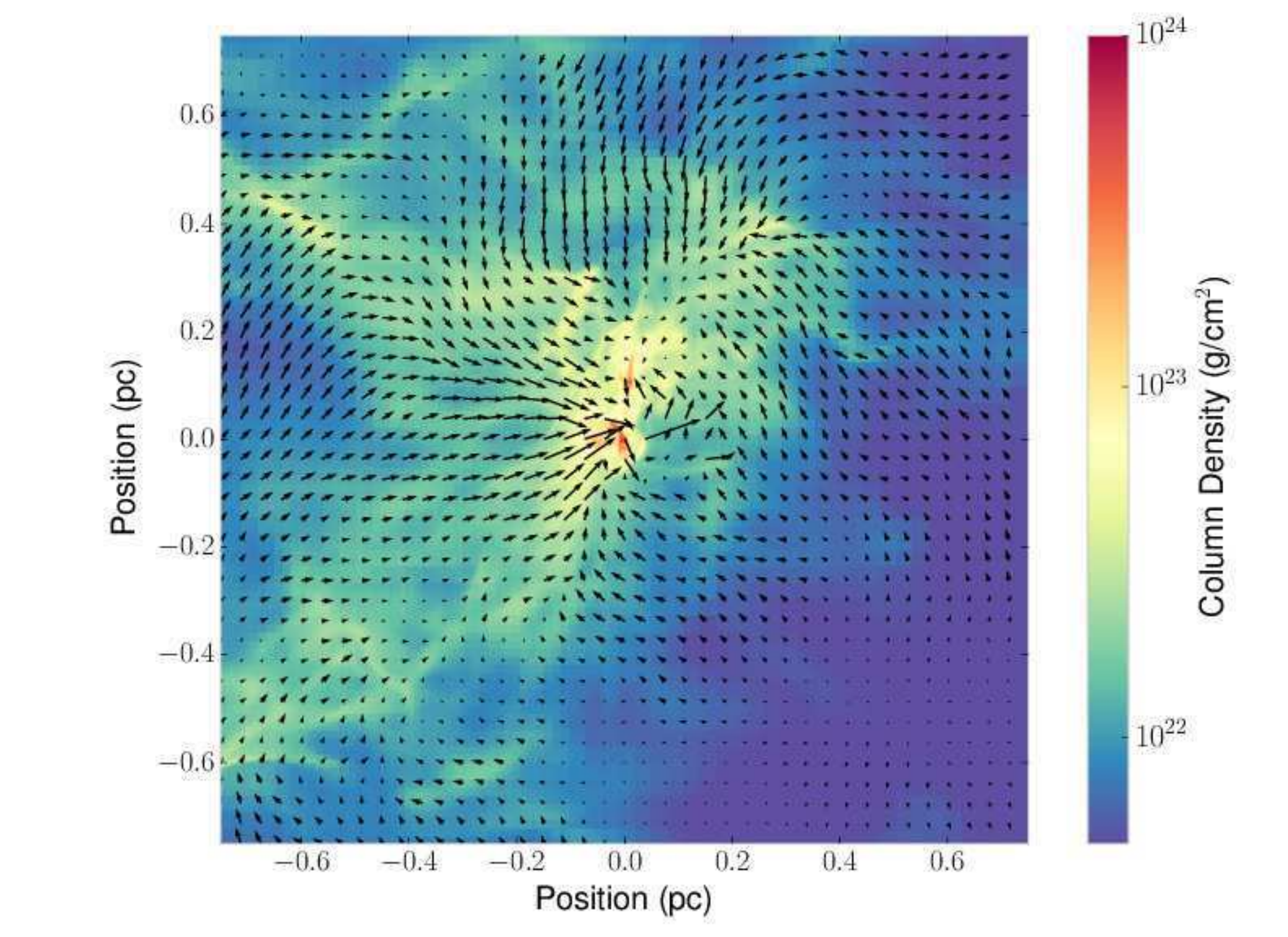}
\caption{A zoom-in of the {\tt MHD1200} cloud, in the same region as that shown in Figure \ref{fig:m1200_large_scale_orientation} (left panel), centred on the highest-density gas and framing a (1.5 pc)$^2$ region. Column density is shown in colour with density-weighted projected velocity vectors overplotted, which emphasizes the motion of the higher-density gas. The initial \alfven Mach number is 0.99, and the initial virial parameter is $\alpha_{\mathrm{vir}} = 0.56$, i.e.~the cloud is highly bound and undergoing gravitational collapse.}
\label{fig:m1200_projected_velocities}
\end{figure*}

\begin{figure*}
\includegraphics[width=0.7\textwidth]{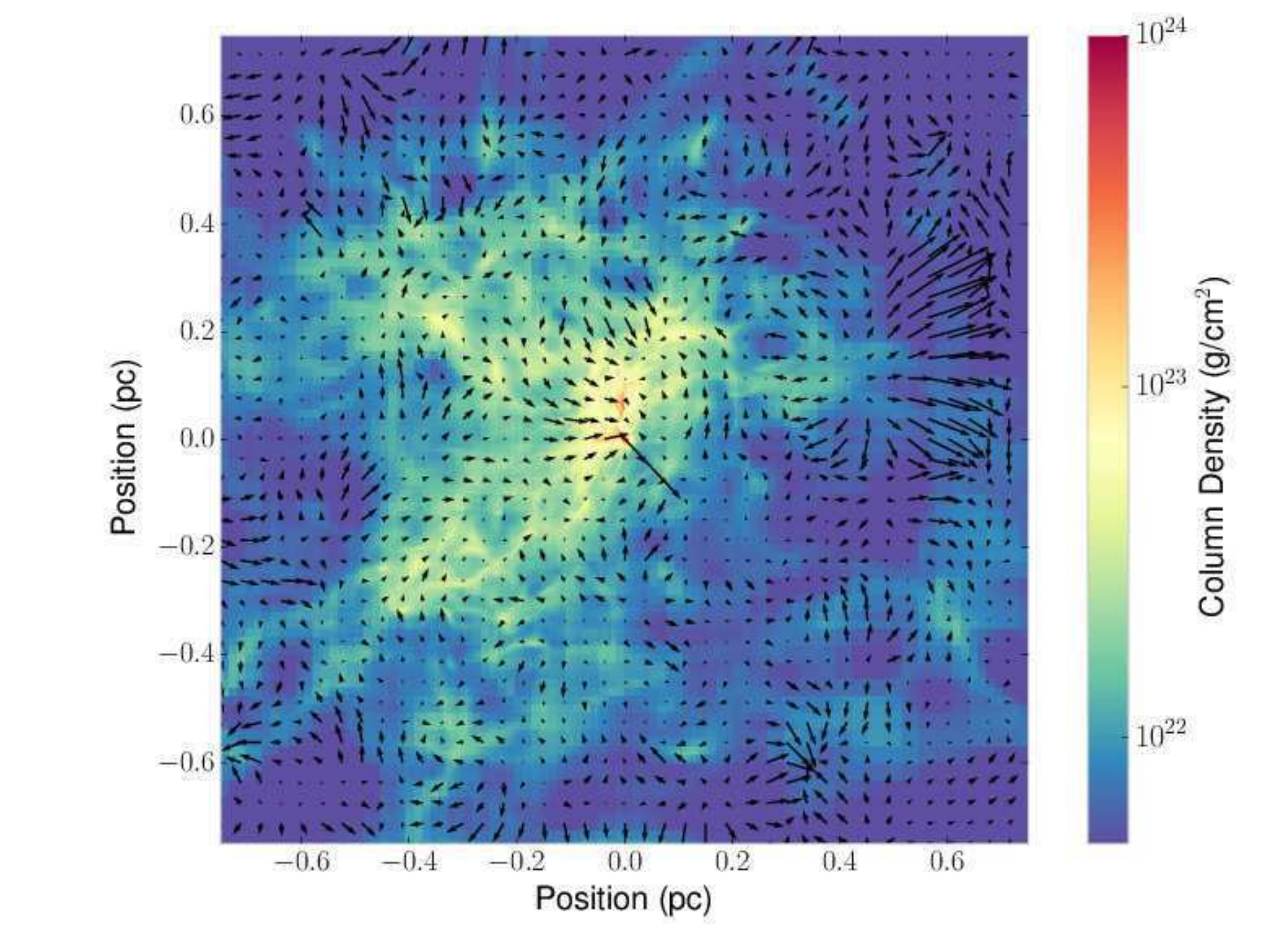}
\caption{A zoom-in of the {\tt MHD500} cloud, in the same region as that shown in Figure \ref{fig:m500_large_scale_orientation} (left panel), centred on the highest-density gas and framing a (1.5 pc)$^2$ region. Column density is shown in colour with density-weighted projected velocity vectors overplotted, which emphasizes the motion of the higher-density gas. The initial \alfven Mach number is 0.92, slightly lower than the {\tt MHD1200} case in Figure \ref{fig:m1200_projected_velocities}. However, the initial virial parameter is $\alpha_{\mathrm{vir}} = 0.95$, i.e. the turbulent kinetic energy is roughly in equilibrium with the gravitational binding energy.}
\label{fig:m500_projected_velocities}
\end{figure*}

The middle panel of Figure \ref{fig:m1200_large_scale_orientation} shows the autocorrelation of the ``column density'' plot from the left panel after downsampling to a 200x200 pixel greyscale image. The largest modes in the initial turbulent velocity field create a main filament ``trunk'' that is home to some of the highest-density gas in the simulation. In performing an autocorrelation of the column density map, the large-scale structure is also the most similar, and features prominently in the autocorrelation map. We contour several levels, from 1\% (the outermost contour) to 10\% of the peak value. We then fit a linear function to the pixel coordinates inside the 1\% contour, weighing each pixel by its base-10 logarithm value. In this way, we select the orientation of the long axis of the main structure in our molecular cloud clump. This approach is similar to that of \citet{Li+2013}, who used the pixel positions of the contour. We found better results using the pixels interior to a given contour, with a weighting based on their pixel value. 

We are interested in measuring the magnetic field orientation and comparing it to the orientation of cloud clump. We overplot in blue the magnetic field lines on the autocorrelation map. The field lines are based on the magnetic field measurements taken from the left panel. We see that they are still largely oriented vertically, especially in the lower-density regions, but appear to align weakly with main trunk filament and some other high-density branches.

The right panel of Figure \ref{fig:m1200_large_scale_orientation} shows two histograms and two vertical grey lines. The solid right line gives the angle of the ``trunk'' filament from the middle panel, measured relative to vertical, while the dashed left line is offset by $90^\circ$, i.e~perpendicular to the trunk. The histogram in blue is based on all the density-weighted magnetic field averages measured for the left panel. We see that most gas is still oriented vertically, with relatively little deflection ($\pm 10^\circ$) from its original orientation (vertical). However, shown in red is the distribution of magnetic field orientations for gas with a mean density $n > 3 \times 10^5$ cm$^{-3}$ ($\rho \gtrsim 10^{-18}$ g/cm$^3$). This relatively high density gas has somewhat a bimodal distribution, showing a preference for alignment with the main filament trunk (the right peak in the distribution). The smaller left peak in the distribution is towards a perpendicular orientation with the filament (the left vertical grey line), but not exactly perpendicular. Accretion flow onto the main trunk filament, and then along it, has dragged magnetic field lines along with it such that they are deflected towards the filament. This accretion flow is responsible for this second (left) peak in the bimodal distribution. At lower density thresholds, the twin peaks of the bimodal distribution combine in the centre, whereas at higher thresholds, the bimodal distribution persists until too few cells remain above the threshold density for good sampling. 

We now compare this to Figure \ref{fig:m500_large_scale_orientation} (the virialized cloud simulation), which applies the same analysis to the {\tt MHD500} simulation. The analysis is done in the same way, except that for the autocorrelation map contours we plot evenly-spaced levels between 0.1\% and 10\% of the peak value. The {\tt MHD500}, being the tighter, more compact cloud with a higher overal gas density relative to the {\tt MHD1200} simulation, is actually less bound.

The measurements for Figure \ref{fig:m500_large_scale_orientation} are taken at the same number of freefall times as compared to the {\tt MHD1200} simulation to allow for a fair comparison. 

The higher overall gas densities in {\tt MHD500} meant a more strongly peaked central value in the autocorrelation map. For this reason, we contoured down to 0.1\% of the peak value in the middle panel of Figure \ref{fig:m500_large_scale_orientation} to trace more of the overall structure.

\begin{figure*}
\includegraphics[width=1.0\textwidth]{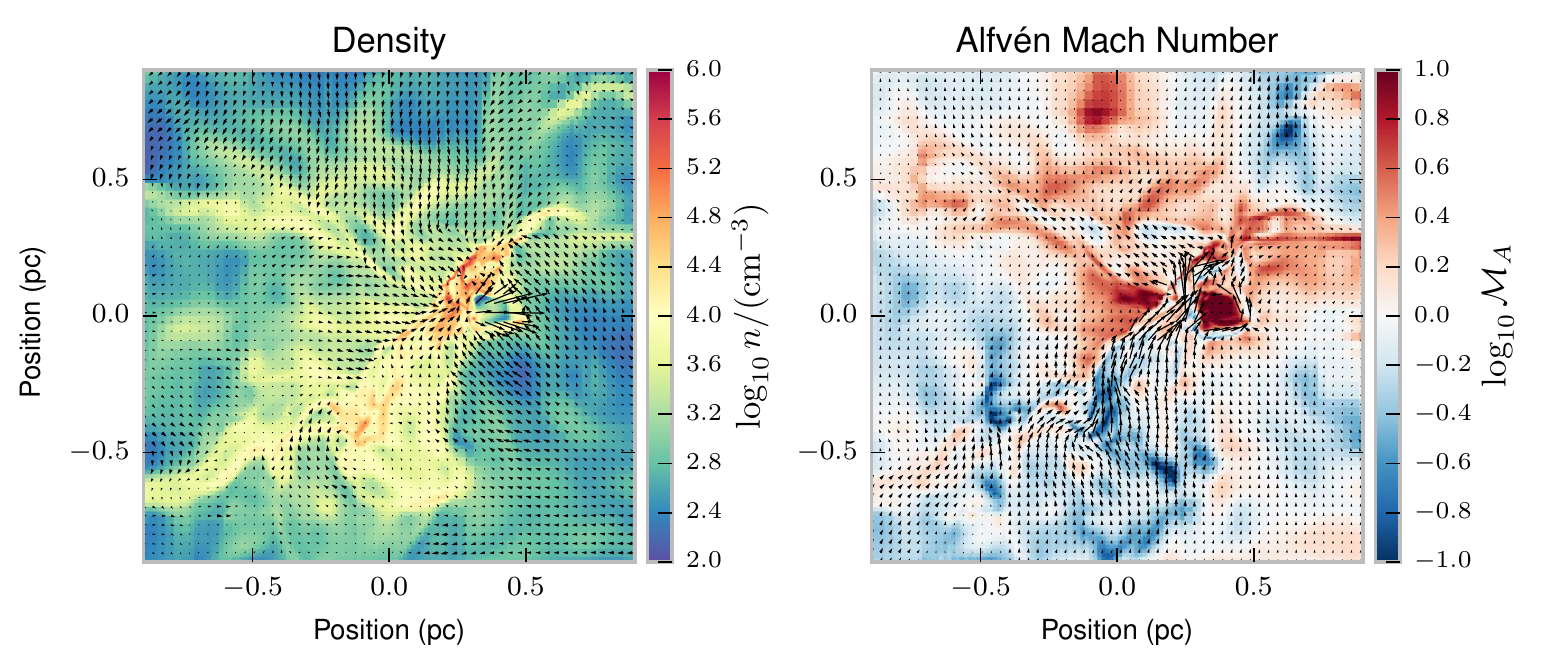}
\caption{{\it Left:} Volume density slice through the molecular cloud of our {\tt MHD1200} simulation after 250 kyr of evolution with arrows indicating the velocity field. The colours cover a range in volume densities from $n = 100$ cm$^{-3}$ to $n = 10^6$ cm$^{-3}$. Photoionization feedback from a cluster of stars (not shown) has begun forming an \HII region at the side of the main trunk filament. {\it Right:} Local \alfven Mach number with arrows indicating the magnetic field. The colours are scaled logarithmically from $\mathcal{M}_A = 10^{-1}$ (blue) to $\mathcal{M}_A = 10^{1}$ (red). White regions have values for the \alfven Mach number of $\mathcal{M}_A \approx 1$, indicating that the turbulent energy is balancing the magnetic energy. Sub-Alfv\'{e}nic regions ($\mathcal{M}_A < 1$) have stronger magnetic fields.}
\label{fig:m1200_slice_dens_alfven}
\end{figure*}

\begin{figure*}
\includegraphics[width=1.0\textwidth]{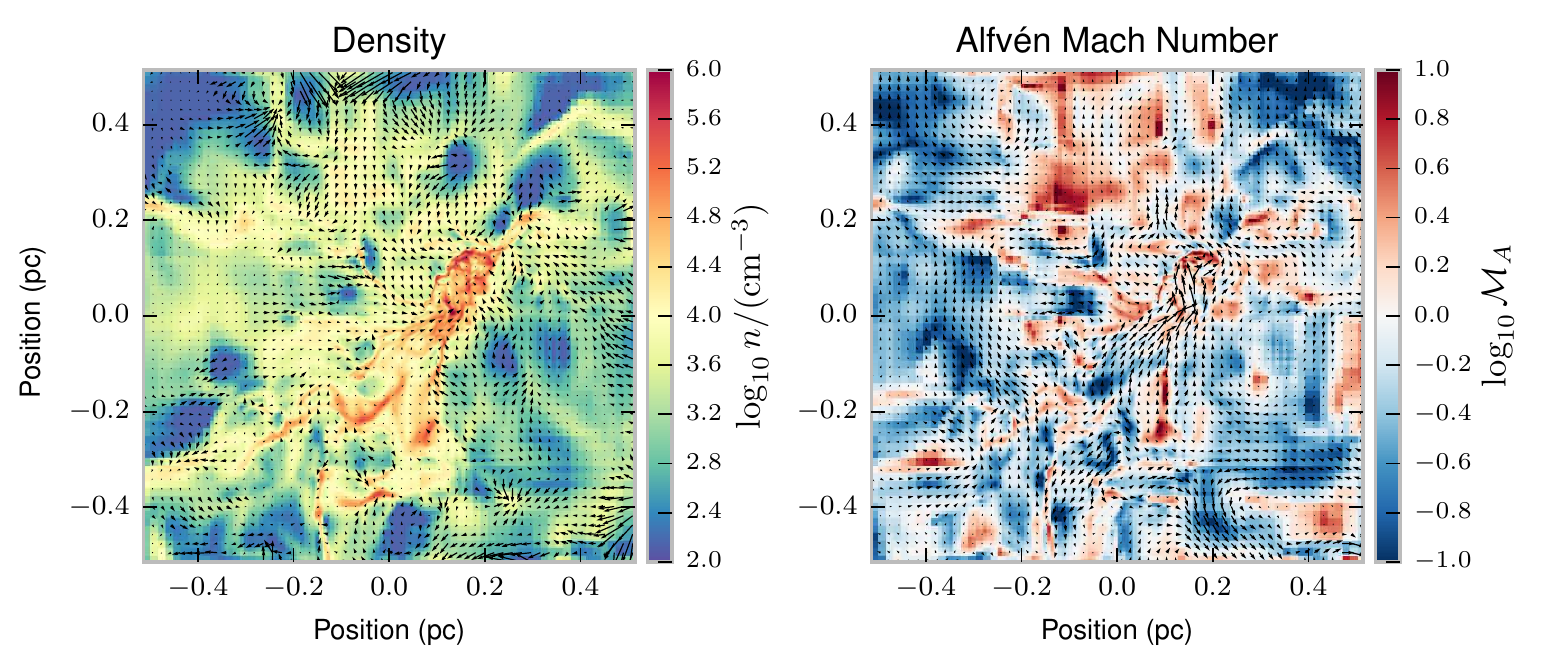}
\caption{The same as in Figure \ref{fig:m1200_slice_dens_alfven}, except for the {\tt MHD500} simulation. As the simulation volume is smaller, the panels have been made proportionally smaller, while still centering on the densest part of the simulation. The snapshot of the simulation is taken after 150 kyr of evolution, which is earlier than in {\tt MHD1200}, but at the same number of freefall times to permit better comparison.}
\label{fig:m500_slice_dens_alfven}
\end{figure*}

The magnetic field orientations appear much more chaotic in Figure \ref{fig:m500_large_scale_orientation}, despite the sound speed and RMS Mach number of the turbulence being the same in both simulations. The initial \alfven Mach number are also virtually identical, suggesting that the magnetic field structure should be similarly ordered or disordered. This is, however, not the case. The difference is entirely on account of the relative boundedness of each cloud.

The virialized cloud in {\tt MHD500} shows no strong preferred ordering of the magnetic field in 2D, as is seen in the right panel of Figure \ref{fig:m500_large_scale_orientation}. Figure \ref{fig:m500_large_scale_orientation} also shows less obvious large-scale structure, as evinced by the autocorrelation map, which appears much more square when compared to Figure \ref{fig:m1200_large_scale_orientation}. This is on account of {\tt MHD500} being in virial equilibrium. {\tt MHD1200}, being the more bound cloud, has undergone strong gravitational collapse onto the central filamentary structure, which appears very prominently in the autocorrelation map. The gravitational collapse has dragged the magnetic field structure with it, which accounts for the more regular structure seen in Figure \ref{fig:m1200_large_scale_orientation} compared to Figure \ref{fig:m500_large_scale_orientation}.

The lower-density gas ($n \leq 3 \times 10^5$ cm$^{-3}$) forms an approximately Gaussian distribution in angle relative to ``North'', centered at 0$^\circ$. This distribution is shown in blue in the right panel of Figure \ref{fig:m500_large_scale_orientation}. The higher-density gas ($n \leq 3 \times 10^5$ cm$^{-3}$), shown in red, is spread over many angles, but, if anything, tends to be oriented more perpendicular to the main trunk filament, an angle indicated by the left vertical gray line. This pattern persists at other threshold densities.

In Figure \ref{fig:m1200_projected_velocities} we show the column density projection of the same region as in the left panel of Figure \ref{fig:m1200_large_scale_orientation} of the {\tt MHD1200} simulation. We then overplot the density-weighted average velocity vectors to show the average flows onto the main filament. We see a pattern of flow both along the long axis of the main trunk filament and accretion onto the filament radially. The flow appears to converge onto the central, densest region. Compare this to Figure \ref{fig:m500_projected_velocities}, which is the equivalent but for the {\tt MHD500} simulation, displaying the same region as in the left panel of Figure \ref{fig:m500_large_scale_orientation}. The density-weighted average velocity shows a pattern of randomly-oriented flows. The difference, again, can be entirely attributed to the degree of boundedness in each case, with {\tt MHD1200} undergoing strong gravitational collapse and {\tt MHD500} exhibiting a relative balance of kinetic and gravitational energies.

To examine the relationship between magnetic field ordering and the \alfven Mach number more closely, we plot volume density slices through the center of our simulations showing the gas structure and compare these to slices of the local \alfven Mach number. We also compare the velocity and magnetic field structure. We show these in Figures \ref{fig:m1200_slice_dens_alfven} and \ref{fig:m500_slice_dens_alfven} for the {\tt MHD1200} (strongly bound) and {\tt MHD500} (virialized) models, respectively.

These two figures were taken at the same number of freefall times, $t_{\textrm{ff}} \approx 0.14$, which corresponds to about 250 kyr in the {\tt MHD1200} simulation and 150 kyr in the {\tt MHD500} simulation. In Figure \ref{fig:m1200_slice_dens_alfven}, we see that the velocity field is channeling mass both onto the main trunk filament and along it. This resembles the mass flows measured for the cluster-forming region in the Serpens South molecular cloud studied in \citet{Kirk+2013}. Material appears to be flowing along the long axis of our main trunk filament, feeding a tight cluster of stars that is driving the \HII region already visible in the left panel of Figure \ref{fig:m1200_slice_dens_alfven}. There is also material flowing onto the main filament radially, and this radial accretion appears stronger than the flows along the filament's long axis, as it was in \citet{Kirk+2013}. The right panel of Figure \ref{fig:m1200_slice_dens_alfven} shows the local \alfven Mach number. More sub-Alfv\'{e}nic regions possess stronger magnetic fields and slower accretion flows. The magnetic field is depicted with arrows in the right panel, showing how the magnetic field has been concentrated into the main trunk filament so that it lies parallel with the filament long axis.

In Figure \ref{fig:m500_slice_dens_alfven} we see the result of trans-Alfv\'{e}nic turbulence in virial balance with the gravitational forces in the {\tt MHD500} simulation. Recall that both simulations were initialized with the same turbulent velocity field, similar mass-to-flux ratios, and the same radial density profile. The {\tt MHD500} simulation is actually at higher average density---a more compact setup. However, after 150 kyr of evolution, the magnetic field has become disordered and the local \alfven Mach number is a patchwork of ripples, alternating islands of sub- and super-Alfv\'{e}nic regions, encircled with magnetic fields that do not possess very much coherent large-scale structure. There is some similarity with the {\tt MHD1200} simulation, however: the field is generally lying parallel to the long axis of the main filament.

The consensus so far is that turbulence plays a major role in trans-Alfv\'{e}nic molecular clouds. In the trans-Alfv\'{e}nic regime, neither turbulence nor magnetic fields have the clear upper hand. Unlike the sub-Alfv\'{e}nic regime, which dominates in the diffuse ISM and where magnetic fields clearly channel flows, filaments are more the result of turbulence and not of slow accretion flow along dominant field lines. The large-scale turbulent modes give rise to the primary filaments---the trunk---via shock compression, and magnetic field lines are dragged along with the gas, pushed together so that they lie together within the main filaments, parallel to their axes.

In our case, wherein turbulence and magnetic forces are nearly in balance, we show that the discriminating factor is likely whether gravity dominates over the other energies. The {\tt MHD1200} simulation was substantially sub-virial ($\alpha_{\textrm{vir}} = 0.56$), whereas magnetic fields were chaotic in the virially-balanced case of {\tt MHD500}.

\subsection{Velocity fields in 2D}

\begin{figure*}
\includegraphics[width=1.0\textwidth]{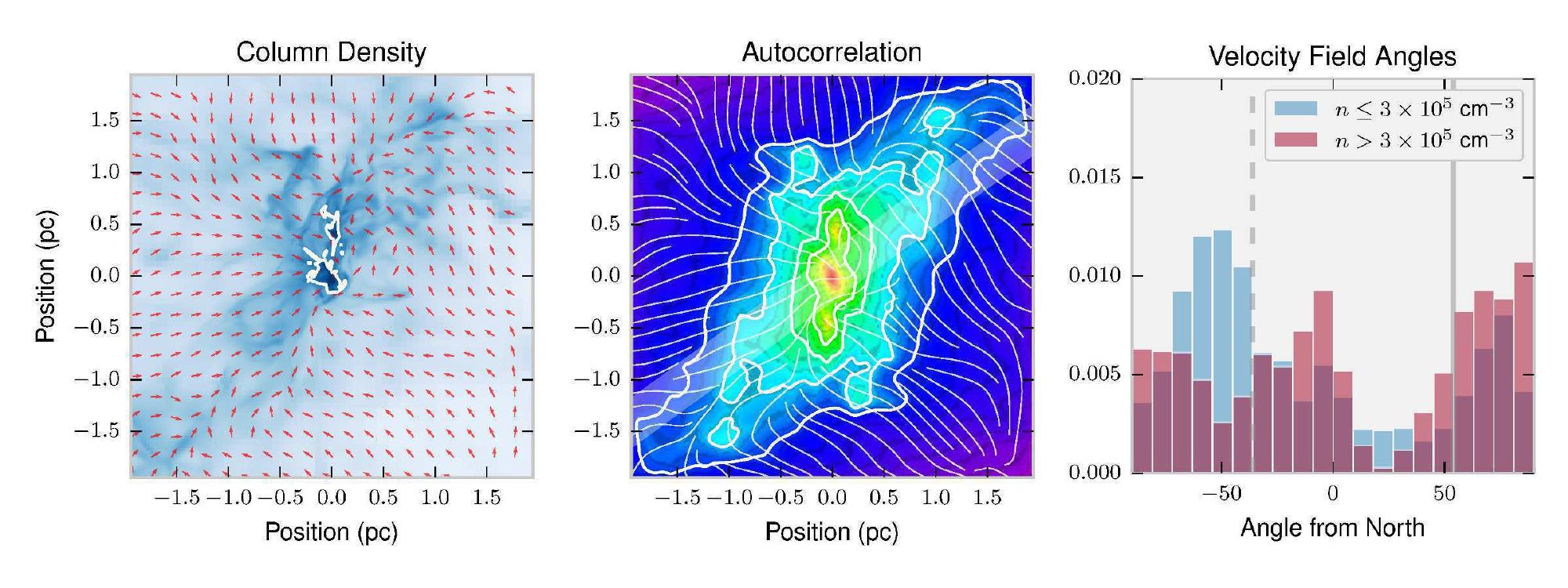}
\caption{{\it Left:} Density projection along the $y$-axis of our {\tt MHD1200} simulation at 250 kyr of evolution with density-weighted projected velocity field overplotted in red arrows. {\it Middle:} The autocorrelation of the previous column density project, which highlight self-similar structure. Contours highlight levels from 1\% to 10\% of peak correlation values. The shaded diagonal bar represents cloud orientation as previously calculated for Figure \ref{fig:m1200_large_scale_orientation}. Velocity streamlines based on the values measured for the left panel are overplotted. {\it Right:} The histogram of the velocity field orientations based on the values measured for the left panel, normalized so that the total area is 1. The orientations are measured relative to ``north''. The blue histogram shows the low-density gas distribution, while red indicates only the orientations of high-density gas. The vertical grey lines indicate the angle of the best fit line to the large-scale structure (solid right line), and the angle offset by 90$^\circ$ (dashed left line).}
\label{fig:m1200_large_scale_velocities}
\end{figure*}

\begin{figure*}
\includegraphics[width=1.0\textwidth]{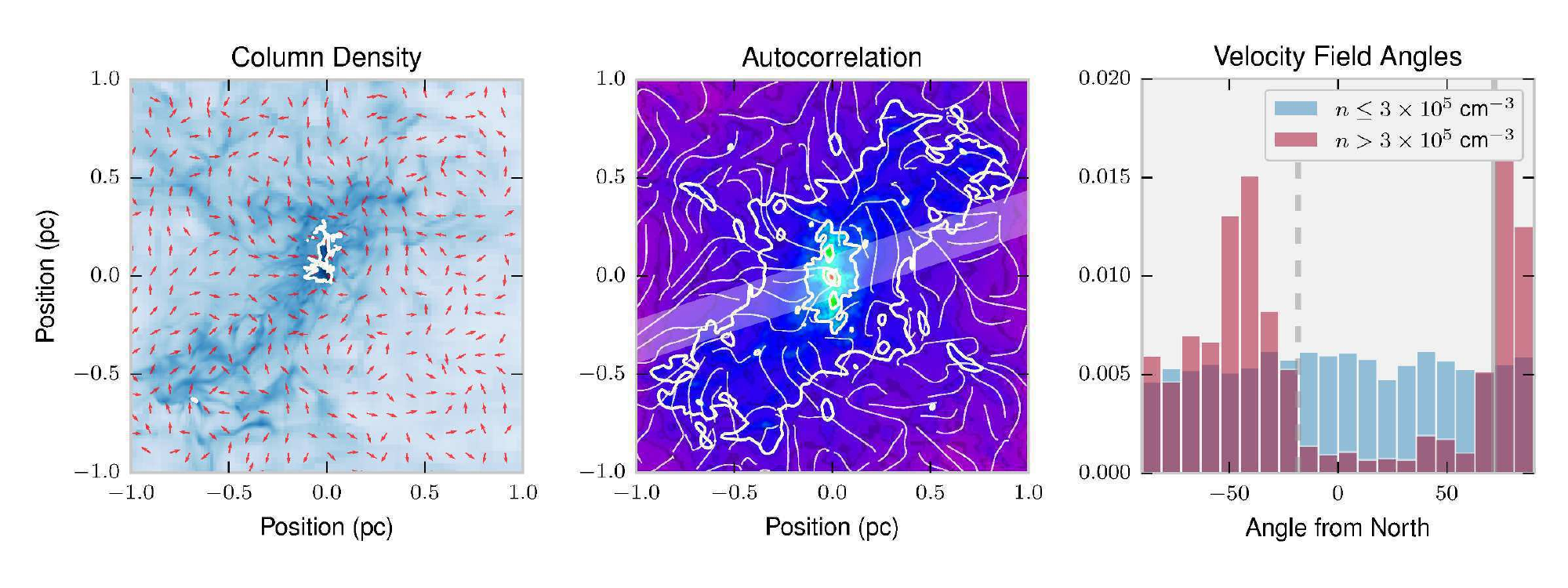}
\caption{The same as in Figure \ref{fig:m1200_large_scale_velocities}, except using {\tt MHD500} at 150 kyr of evolution. Because of the higher average gas density, we draw contours in the middle panel from 0.1\% to 10\% of peak correlation values. The best-fit line is still calculated based on the outermost contour. The data is taken at approximately the same number of freefall times as in the {\tt MHD1200} simulation.}
\label{fig:m500_large_scale_velocities}
\end{figure*}

In Figures \ref{fig:m1200_large_scale_velocities} and \ref{fig:m500_large_scale_velocities}, we show the large-scale velocity patterns in projection for the {\tt MHD1200} and {\tt MHD500} simulations, respectively. In each of these figures, the left panel shows the mean projected gas density in blue. Overplotted are velocity vectors computed by taking the density-averaged mean velocity along the line of sight. The white contour is the threshold density used in the right panel to separate ``low'' and ``high'' density gas.

The middle panels compute the autocorrelation of the mean gas density, just as we had done previously for Figures \ref{fig:m1200_large_scale_orientation} and \ref{fig:m500_large_scale_orientation}. The only difference in this case is that the overplotted streamlines reflect the velocity structure instead of the magnetic field structure.

Finally, the right panels in Figures \ref{fig:m1200_large_scale_velocities} and \ref{fig:m500_large_scale_velocities} show the histograms of the velocity field orientation angles, relative to ``north'' (the $z$-axis), based on the angles in every pixel of the left panel. The data is divided into ``low-density'' and ``high-density'' pixels, based on the mean gas density along the line of sight, with $\rho = 10^{-18}$ g/cm$^3$ ($n = 2.8 \times 10^5$ cm$^{-3}$) as the threshold. The high-density regions are contoured in the left panel of each figure, and show a dense core region in each case.

The velocity angle histogram in Figure \ref{fig:m1200_large_scale_velocities} shows a bimodal distribution for the low-density gas in blue for the {\tt MHD1200} simulation. These do not lie perfectly parallel or perpendicular to the main trunk filament angles (the right and left vertical grey lines, respectively), but do follow a general pattern: most of the low-density gas is flowing onto main trunk filament at an angle roughly perpendicular to it, with a smaller fraction of the gas also flowing along the main filament.

The high-density core region shows a high degree of flow along the main filament with another significant portion coming in laterally. This is shown by the red histogram.

In the {\tt MHD500} simulation, the histogram of gas velocities shows a different picture. The low-density gas is randomly oriented. There are no preferred accretion channels and gas appears to flow in all directions. This is in strong contrast to the high-density gas of the inner core region (shown in red), which shows extremely strong flow along the main trunk filament, and roughly perpendicular to it. This region is likely undergoing gravitational collapse and the main filament appears to set up accretion pathways, either along and perpendicular to it, to supply this central region with gas. 

\subsection{Magnetic field orientation relative to filaments in 3D}\label{sec:small_scale_structure}

What is happening to the relative orientation of the filaments and the magnetic field lines at the level local to the filaments? Using the filamentary structure extracted from the 3D data cubes by \disperse, we analyze the filament spines by visiting the vertices and locally measuring the tangent vector, the magnetic field vector, and other physical variables such as the mass density. This procedure allows us to construct histograms of the relative orientation of the magnetic field and the filament orientation. 

We take the relative angle $\theta$ of the filament tangent vector and magnetic field vector. Some authors measure $\cos\theta$ or $\cos(2\theta)$, but trigonometric projection can cause histograms to suggest a strong trend toward $\cos\theta = \pm 1$. (Imagine a unit circle evenly sampled in the angular dimension. A histogram of the cosine of these angles will collect more samples in the bins nearest $\cos \theta = \pm 1$ unless the bin width is carefully modified as a function of $\theta$.) To avoid these problems, we produce histograms in relative angle only.

We are also careful to compare to the case of random orientation. Because of the two angular degrees of freedom in three-dimensional space, a histogram of relative orientation will have a higher proportion of vectors with angles closer to perpendicular than parallel. Any true tendency in nature towards a parallel or perpendicular orientation of filaments relative to magnetic fields should therefore be measured relative to the case of random orientation.

\begin{figure*}
\includegraphics[width=0.8\textwidth]{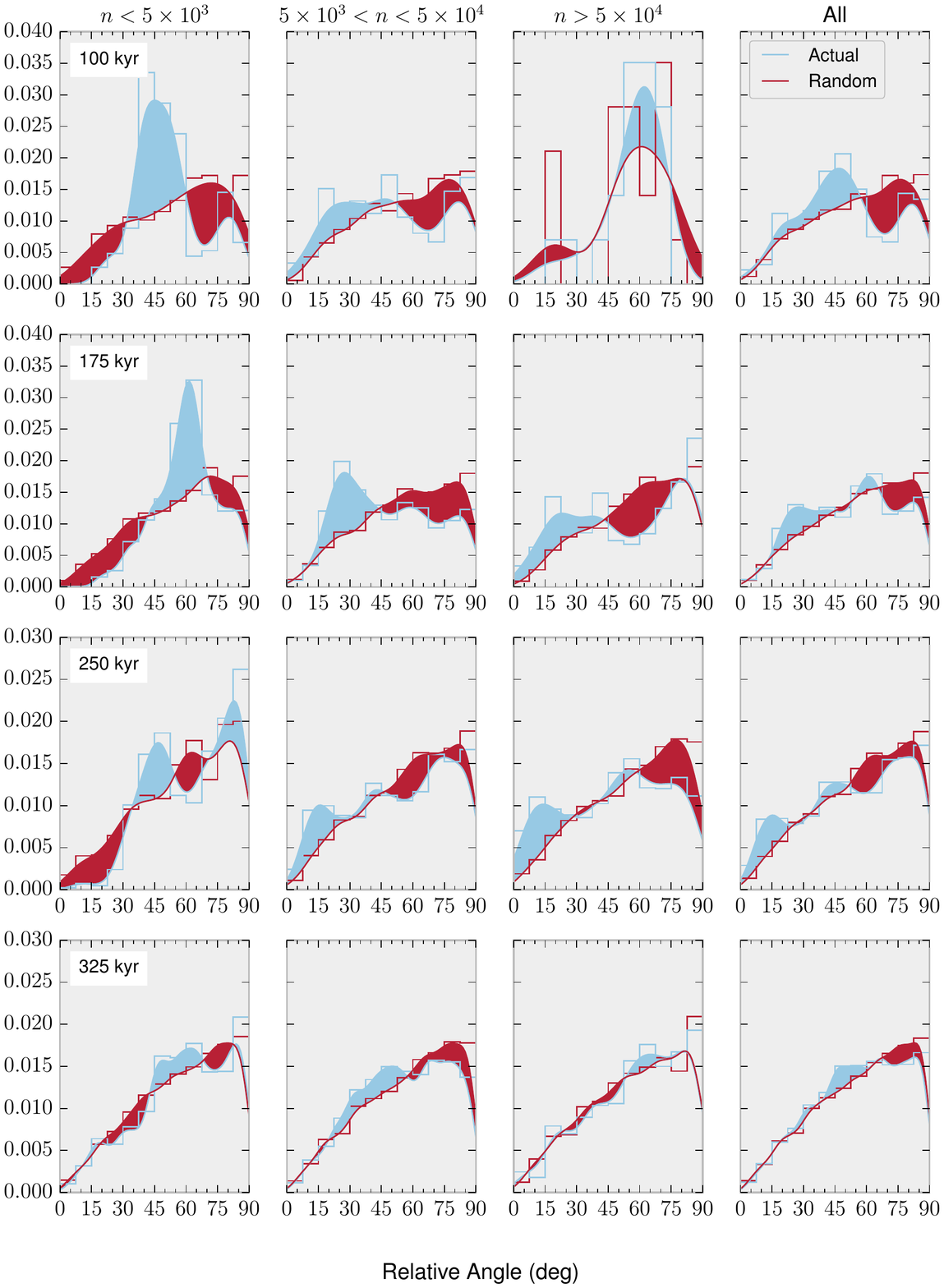}
\caption{Sequence of histograms from the {\tt MHD1200} simulation. The area under each curve has been normalized to 1. By tracing along the filaments in 3D through the simulated volume, we produce histograms of the relative orientation of the magnetic field and the filament. In each panel, we compare the relative orientation measured in the data (blue) with the histogram that would have resulted if the magnetic field had been randomly oriented (red). In addition to a standard step-shaped histogram, a kernel density estimate (KDE), with a Gaussian kernel, has been run over the data and is shown via the smooth curves, providing a continuous analog to the discretely-binned histogram data. Shaded areas indicate either an exceess relative to random (blue) or a deficit (red). Each row of panels gives the state of the simulation at the indicated time. Columns restrict the relative orientation data to the indicated density regims, as measured locally along the filament spine. The last column applies no density selection.}
\label{fig:m1200_rel_orientation_hist}
\end{figure*}

For our {\tt MHD1200} simulation, for the series of time points under consideration, which were selected to be spaced roughly evenly throughout the simulation, we partition the data into three density groups: ``low'' ($n / \textrm{cm}^{-3} < 5\times10^3$), ``medium'' ($5\times10^3 < n / \textrm{cm}^{-3} < 5\times10^4$), and ``high'' ($n / \textrm{cm}^{-3} > 5\times10^4$). These are the first three columns shown in Figure \ref{fig:m1200_rel_orientation_hist}. The final column gives the histogram for all data, making no density cuts. In this figure, the rows represent the time evolution of these histograms of relative 3D orientation. The times at which these histograms were taken are indicated the top-left corner of the first panel in each row. 

In each panel, two histograms are shown. In blue is the distribution of relative angles presented by our data---the ``Actual'' case. We compare the measured relative angles to the case of random orientation, shown in red and labeled ``Random''. In this case, the local magnetic field is assigned a random orientation in 3D space and the relative orientation to the filament is measured. The sample size for the ``Random'' case therefore matches the ``Actual'' case.

\begin{figure*}
\includegraphics[width=0.8\textwidth]{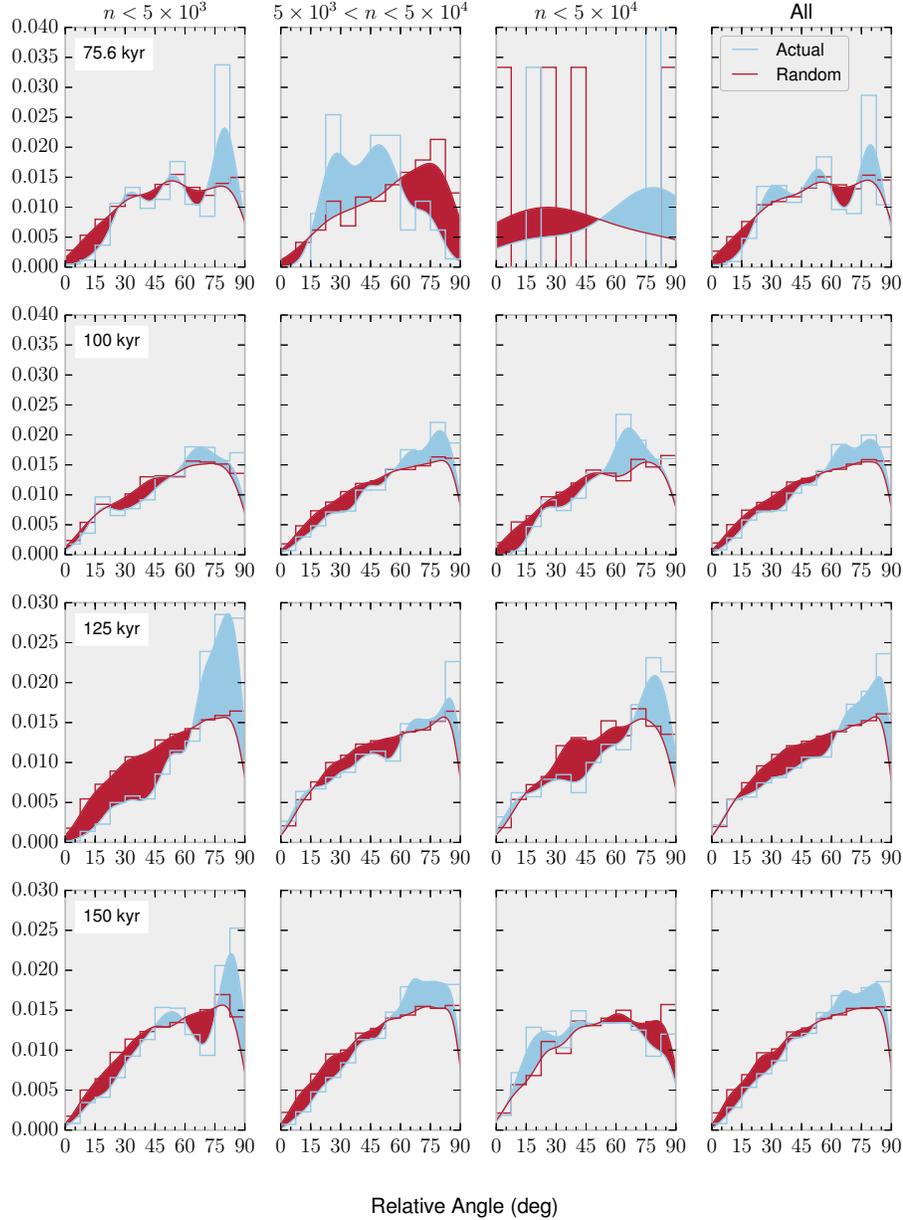}
\caption{The same as in Figure \ref{fig:m1200_rel_orientation_hist}, except for our {\tt MHD500} simulation.}
\label{fig:m500_rel_orientation_hist}
\end{figure*}

In addition to performing the standard step-shaped histogram with 12 bins over 90$^\circ$ of angle, we apply a kernel density estimate (KDE) using Gaussian kernels. KDEs are a continuous analog to the traditional histogram with its discrete number of bins. Figures \ref{fig:m1200_rel_orientation_hist} and \ref{fig:m500_rel_orientation_hist} show both representations. Additionally, we shade any ``excess'' (blue) or ``deficit'' (red) relative to the ``Random''.

In studying Figure \ref{fig:m1200_rel_orientation_hist}, the {\tt MHD1200} simulation, a few trends appear. Although containing more mass overall, the volume of the {\tt MHD1200} is also much larger, causing the mean gas density in the {\tt MHD1200} simulation to be about one third the value in the {\tt MHD500} simulation. We see that in the ``All''-density column there is a deficit in the distribution of relative angles around 90$^\circ$. This pattern shows up in many of the panels where the data has been further segmented based on underlying mass density.

In the low-density column, where we measure relative angles at filaments segments situated in gas at densities $n < 5\times10^3$ cm$^{-3}$, there is neither a trend towards parallel nor perpendicular orientation. Instead, there is an excess relative to random at angles between 30$^\circ$ and 75$^\circ$. This low-density regime represents the outer regions of our simulation volume. The magnetic fields lines, initially coherent and oriented parallel to the $z$-axis, have been dragged inward during the molecular cloud clump's gravitational collapse. In the outer regions of the simulation volume, this inward dragging may account for the observed excess at ``middle'' angles.

At relatively early times, after 100 kyr of evolution (first row of Figure \ref{fig:m1200_rel_orientation_hist}), most of the gas is still at low density. It has not had time to collapse to higher density, i.e.~into the main trunk filament. In the low-density gas histogram, the peak is measured around $45^\circ$ relative orientation between filaments and B-field. This may be on account of the main filament trunk having this approximately this orientation, which may just be starting to form as a result of the largest turbulent modes. At this time, there is very little gas with densities above $5\times10^3$ cm$^{-3}$.

As the simulation progresses, we observe increased power at near-parallel orientations. See, especially, the ``intermediate'' and ``high''-density panels at 250 kyr of evolution, with corresponding deficits at near-perpendicular orientations. An interesting thing happens between 250 kyr and 325 kyr. Between these times, star formation has occurred in this simulation, and radiative feedback has injected energy into the molecular cloud clump. The measured relative orientations between the magnetic field and the filamentary structure in 3D is much closer to random than in any of the other panels. We attribute this to the energy injected through radiative feedback from high-mass stars, an effect we discuss further in section \ref{sec:radiative_feedback}. Radiative feedback is already underway by 250 kyr, but by 325 kyr has largely disrupted the main filament trunk. The radiative feedback is in the form of ionizing radiation from a tight cluster of massive stars that lead to the formation of an expanding \HII region.

At 175 and 250 kyr, we see a feature indicating parallel relative orientation of the filamentary structure and the magnetic field. It is seen in the medium and high density panels, and also appears more modestly in the ``all data'' panel on the right. As the simulation progress, this feature is apparently washed out, because the alignment isn't seen in the last row. Again, this is on account of radiative feedback from star formation. 

It is important to note that \disperse was run on each simulation plotfile separately, and the filamentary structure extracted from each will be different. The persistence and noise thresholds were held the same for the sake of consistently, but the simulation evolves over time and central regions become denser due to gravitational infall. Therefore the skeletons that \disperse extracts from the set of plot files, though they come from the same simulation, may not necessarily strongly resemble each other. This is in contrast to what was done in \citet{Kirk+2015}, wherein the filament skeletons were found in 2D column density projections and the skeletons kept consistent between time slices using by-hand adjustments as necessary. Owing to the considerably higher complexity of doing this with 3D skeletons and having sparser time sampling, we did not attempt to map out the exact same filaments at each time slice.

Nevertheless, the filament skeletons extracted from one column density projection to the next were largely similar, and \citet{Kirk+2015} demonstrated that the properties of our {\sc flash}-simulated filaments matched those of filaments observed in nature.

In Figure \ref{fig:m500_rel_orientation_hist} we show the same analysis performed on our {\tt MHD500} simulation. {\tt MHD1200} and {\tt MHD500} both have the same angular rotation rate, similar ratios of rotational kinetic energy to gravitational energy, the same initial temperature and RMS Mach number turbulence. We initialized both from the same turbulent velocity field, hence they would have developed similar initial structure. The difference is that {\tt MHD500} is tighter and more compact, with a higher average gas density ($\bar{n}_{500} \approx 1.2\times10^3$ vs $\bar{n}_{1200} \approx 3.9 \times 10^2$ cm$^{-3}$).

At early times it is difficult to discern any pattern in the relative orientations. Most of the gas has not yet collapsed to very high density, and the orientations appear relatively close to random distributed.

This changes as the simulation evolves. In the panels at 100 kyr and 125 kyr, the data shows a tendency towards a perpendicular relative orientation. This is visible even in the low-density panels where $n < 5\times10^3$ cm$^{-3}$. At late times (150 kyr, final row), however, this trend is less clear. The high-density ($n > 5\times10^4$ cm$^{-3}$) at 150 kyr appears either to have reversed the trend, or else the orientations have simply become closer to random. In Figure \ref{fig:vir_m2f_evo}, we showed how at 150 kyr the {\tt MHD500} simulation becomes magnetically critical, meaning that the magnetic flux is sufficient to support against further gravitational collapse. We also showed how the virial parameter, which measures the ratio of kinetic to graviational energy increases sharply toward 150 kyr. We interpret this as gravitational energy being converted to kinetic energy during gravitational collapse, in particular as the cloud clump becomes magnetically critical.

Whereas in the {\tt MHD1200} simulation we saw kinetic energy being injected in the form of radiative feedback from massive stars, in the {\tt MHD500} simulation, in which we did not simulate star formation, gravitational energy is converted to kinetic energy, with similar results: the relative orientation of the magnetic field to the filamentary structure becomes closer to random with the injection of kinetic energy.

\subsection{Virial parameter and mass-to-flux ratio}\label{sec:virial_parameter}

\begin{figure}
\includegraphics[width=88mm]{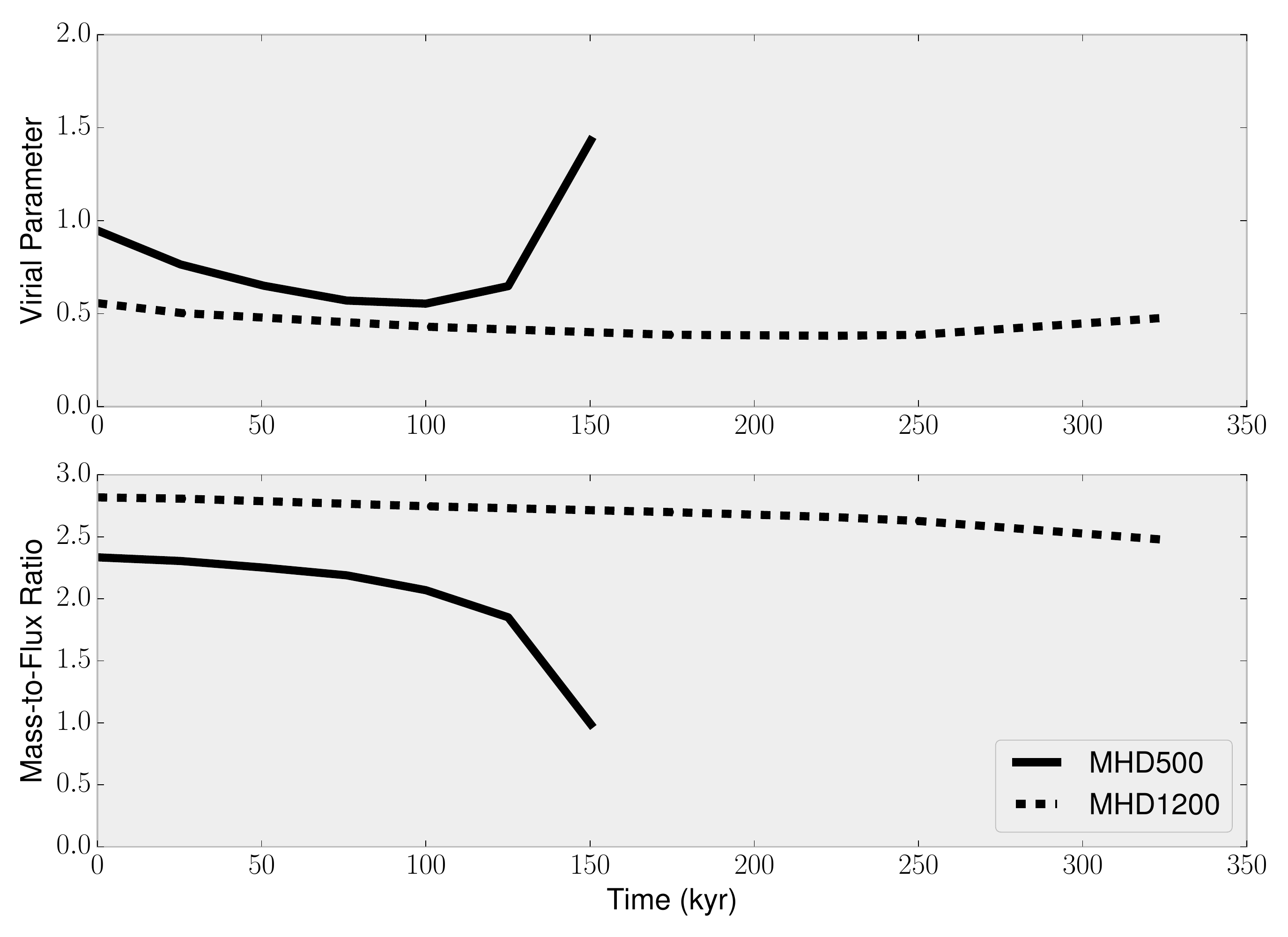}
\caption{The evolution of the virial parameter and the mass-to-flux ratio of our two simulations. These parameters remain fairly steady throughout the {\tt MHD1200} simulation, but the {\tt MHD500} becomes magnetically critical during gravitational collapse. The added magnetic support causes some of the gas in the outer regions to rebound, adding kinetic energy.} 
\label{fig:vir_m2f_evo}
\end{figure}

%
%

We plot the evolution of both the virial parameter and the mass-to-flux ratio in Figure \ref{fig:vir_m2f_evo}. Both are calculated as their volumetric averages with the simulation, and the mass-to-flux ratio is normalized to the critical mass-to-flux ratio (Equation \ref{eqn:mass_to_flux}).

The {\tt MHD1200} simulation begins substantially sub-virial, meaning that as an unmagnetised cloud clump, it would be highly bound and undergo gravitational collapse. Since its mass-to-flux ratio is also supercritical, it does indeed undergo gravitational collapse.

The value of the virial parameter remains relatively level throughout the simulation, decreasing slightly over time as the cloud collapses, but then beginning to increase against around 250 kyr, the time at which massive star formation has resulted in kinetic energy being injected into the simulation via radiative feedback. The mass-to-flux ratio decreases monotonically throughout the simulation as a result of magnetic field lines being dragged slowly inward, following the gravitational collapse.

The {\tt MHD500} simulation behaves rather differently. The simulation evolves quickly, and the virial parameter goes from marginally bound (0.95) to 0.55 after 100 kyr, following which it increases dramatically to 1.43 after 150 kyr. This trend is attributable to what is happening with the mass-to-flux ratio. The simulation begins slightly more magnetised than the {\tt MHD1200} simulation. Magnetic field lines are dragged inward following the gravitational collapse of the gas (which is, on average, 3 times denser than that of the {\tt MHD1200} simulation). The boundary conditions assume that the simulation volume resides in an ambient medium with the same initial magnetic flux desnity. Hence, during gravitational collapse, magnetic field lines are effectively dragged into the simulation volume during gravitational collapse.

The {\tt MHD500} simulation becomes magnetically critical after 150 kyr, meaning that magnetic support is able to prevent further collapse. Gravitational energy gets converted to kinetic energy as the collapse is halted. This increase in kinetic energy is reflected in the virial parameter.

\section{Star formation and the effect of radiative feedback}\label{sec:radiative_feedback}

How does radiative feedback from a forming massive star (which we find in the {\tt MHD1200} simulation) affect the filamentary structure of the cloud? In particular, can it destroy or alter the filamentary accretion flow so as to shut down accretion onto the massive star?

To answer these questions, we ran our {\tt MHD1200} simulation including radiative feedback from star formation. We use a characteristics-based raytracer coupled to a simplified version of the DORIC radiative cooling, heating, and ionization package \citep{FrankMellema1994,MellemaLundqvist2002,Rijkhorst2006} implemented in FLASH for the study of star formation and \HII regions \citep{Peters2010a}.

The {\tt MHD1200} forms a cluster of massive stars near the center of the simulation volume, directly within the main trunk filament. \citet{Schneider+2012,Peretto+2013} observe that the intersections of filaments are the sites of clustered and massive star formation. The main trunk filament in our simulation shows various smaller branches connecting with it. The cluster of massive protostars that forms in our simulation eventually becomes luminous enough to begin ionizing the gas around it and form a \HII region with the appearance of a blister on the side of main trunk filament, similar to the Cocoon nebula in IC 5416 \citep{Arzoumanian+2011}, although the Cocoon nebula is an \HII region powered by only a single B star.

The formation of the \HII region ultimately begins to disrupt and destroy the main filament, creating an expanding cavity of hot ($10^4$ K) gas and injecting a lot of kinetic energy. The \HII region, driven mainly by a single massive star that grows to $16 \Msun$, comes to envelope the entire star cluster and shuts off accretion onto every star.

\begin{figure}
\includegraphics[width=88mm]{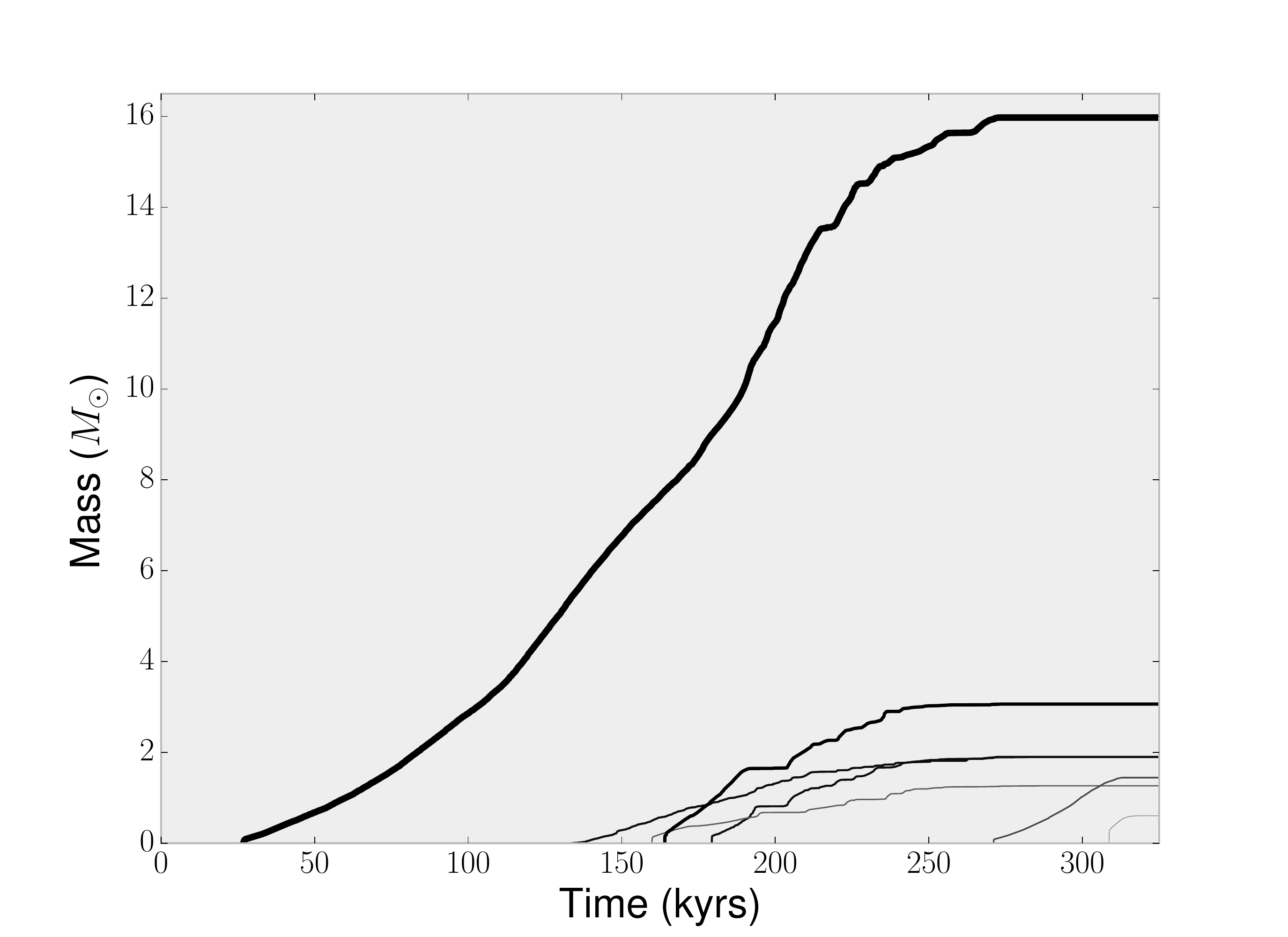}
\caption{The evolution of the sink particles formed in the {\tt MHD1200} simulation. 7 particles are formed near the center of the simulation volume inside the main trunk filament and make up a tight cluster. These accrete mass, the largest of which reaches nearly 16 $\Msun$.}
\label{fig:m1200_sink_masses}
\end{figure}

\begin{figure*}
\includegraphics[width=1.0\textwidth]{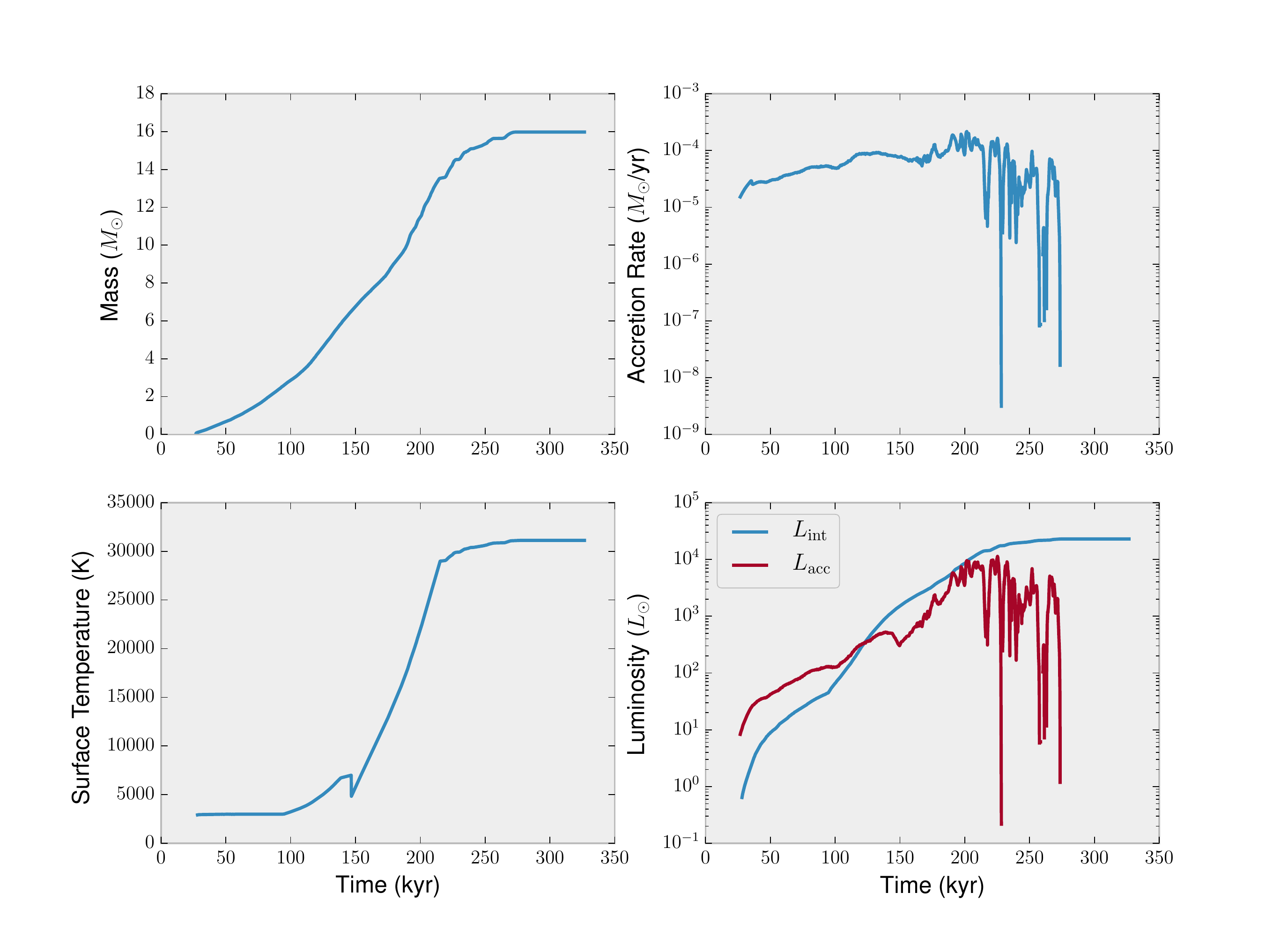}
\caption{Evolution of the most massive star formed as part of the {\tt MHD1200} simulation, which reaches a maximum mass of 16 $\Msun$. The top-left panel shows the history of the mass of the sink particle representing this star. The top-right panel shows the evolution of the accretion rate in units of $\Msun$/yr. The bottom-left panel is the history of the effective (surface) temperature, as computed by our protostellar model. The bottom-right panel shows the intrinsic luminosity and the accretion luminosity of the star.}
\label{fig:m1200_sink_evolution}
\end{figure*}

Figure \ref{fig:m1200_sink_masses} shows the mass evolution of the stars formed in the {\tt MHD1200} simulation. A total of 7 stars are formed, although only one of them becomes a massive star, achieving about 16 $\Msun$. Of the others, one reaches $3 \Msun$, while the others remain below $2 \Msun$. The luminosity of the massive star dominates all the others, achieving a total luminosity of $L_{\textrm{tot}} = 22942 \Lsun$. This star powers the formation of the \HII region. The 10,000 K gas within this region supplies the thermal pressure needed to drive the outward expansion of the bubble and form the blister that eventually begins to disrupt the main filament. A star with a mass of 16 $\Msun$, in a cluster of 7 stars with a total mass of 26.16 $\Msun$ is at the high end---but still observed---range for embedded star clusters \citep{Weidner+2010}.

We take a closer look at the evolution of the most massive star in Figure \ref{fig:m1200_sink_evolution}, which plots the evolution of several of its properties. The top-left panel shows the mass evolution as in Figure \ref{fig:m1200_sink_masses}, but only for the most massive star. After about 250 kyr of evolution, the mass of the star begins to plateau as it envelopes itself in the \HII region of its own making. Less gas reaches the star and its accretion is shut off. We see this play out in the top-right panel of the figure, which shows the accretion rate reach a maximum of around 1--2 $\times 10^{-4}$ $\Msun$/yr, but then begins to drop off. After 275 kyr, the accretion rate has shut off completely and the star ceases to grow. In this panel, we have applied a small amount of smoothing via a moving average filter. 

In the bottom-left panel of Figure \ref{fig:m1200_sink_evolution} we plot the evolution of the effective (surface) temperature of the star. This is computed via a protostellar evolution model that we first implemented in \FLASH in \citet{Klassen+2012a} and used in the study of \HII region variability in \citet{Klassen+2012b}. In particular, the notch seen in the surface temperature near 150 kyr is due to a change in protostellar structure as the star's radius swells and the surface cools temporarily. From 150 to 225 kyr, the surface temperature increases dramatically, which results in a high flux of ionizing photons. The \HII region starts to form during this time as the massive star begins to ionize the gas in its vicinity.

The bottom-right panel of Figure \ref{fig:m1200_sink_evolution} shows the evolution of star's intrinsic luminosity (from nuclear burning or Kelvin-Helmholtz contraction during the earliest phases) and the accretion luminosity. Initially, accretion is the dominant luminosity, but is overtaken by the star's intrinsic luminosity after about 125 kyr of evolution, when the star is between 4 and 6 $\Msun$. We see that the accretion luminosity shuts off completely around 275 kyr.

\begin{figure*}
\includegraphics[width=1.0\textwidth]{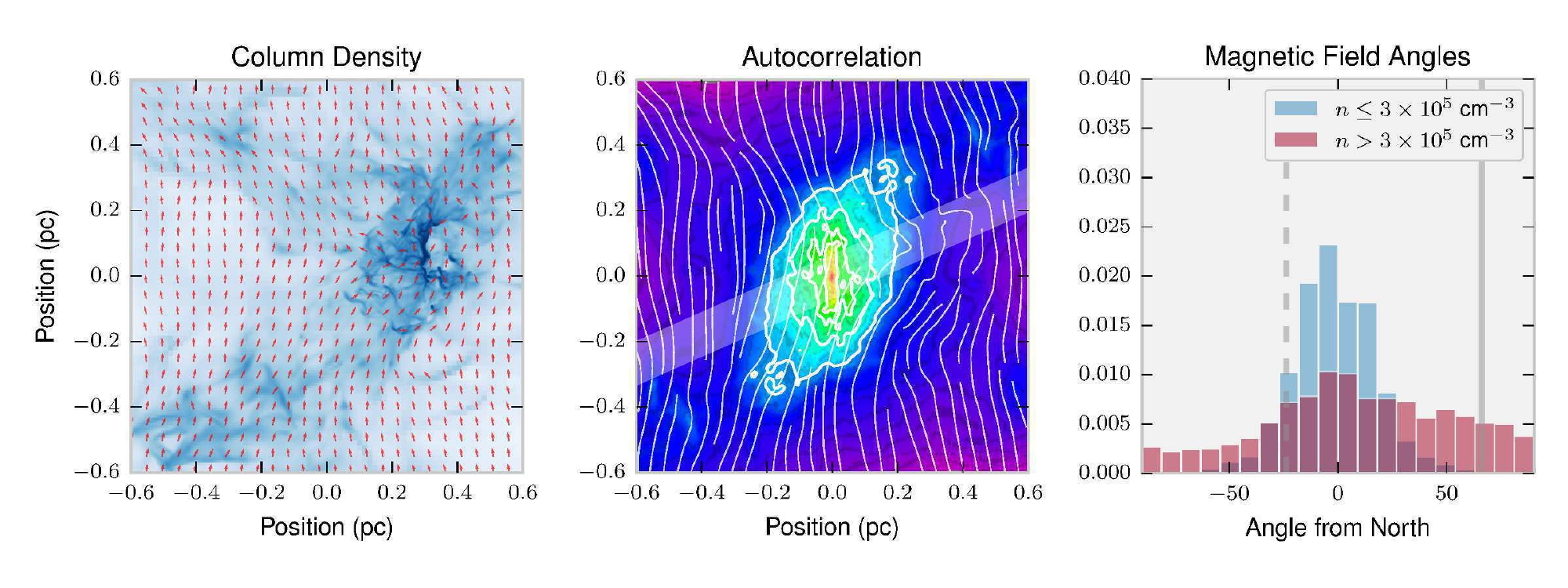}
\caption{The same as in Figure \ref{fig:m1200_large_scale_orientation}, except for the final state of the {\tt MHD1200} simulation at 325 kyr of evolution. Photoionization feedback from a massive star of $\approx 16 \Msun$ has created an expanding \HII region.}
\label{fig:m1200_large_scale_orientation_last}
\end{figure*}

\begin{figure*}
\includegraphics[width=1.0\textwidth]{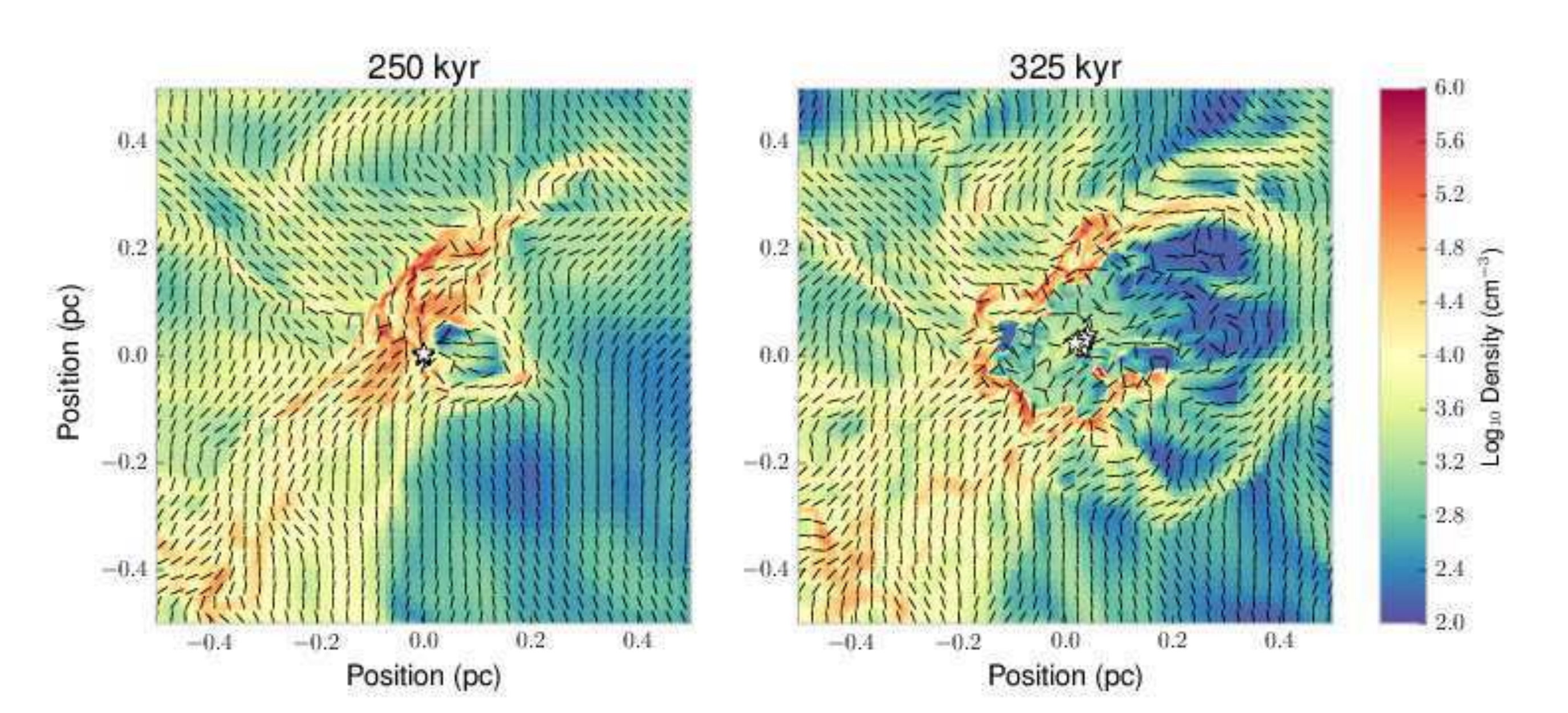}
\caption{Volume density slices through the center of the {\tt MHD1200} simulation volume showing an expanding \HII region as a result of ionizing feedback from the cluster of stars. A single massive star of nearly 16 $\Msun$ dominates all the others and has a luminosity of nearly 23,000 $\Lsun$. This drives a bubble of hot (10,000 K) ionized gas, forming a blister on the side of the main trunk filament. The left panel shows the state of this bubble after 250 kyr of evolution, while the right panel shows the state of the bubble after 325 kyr, near the very end of the simulation. Overplotted on each are magnetic field vectors based on the magnetic field orientation in the plane of the slice.}
\label{fig:m1200_bubble}
\end{figure*}

We repeat the column density autocorrelation analysis in Figure \ref{fig:m1200_large_scale_orientation_last}, this time for the terminal plotfile at 325 kyr of evolution. We zoom in on the inner (1.2 pc)$^2$ of the simulation. The effects of the forming \HII region is seen in the column density plot as a blister on the side of the main filament. Photoionizing feedback injects a large amount of kinetic energy---the gas inside the \HII region is $10^4$ K. The orientation of the magnetic field vectors for the high-density gas ($n > 2.8 \times 10^{5}$ cm$^{-3}$) is almost random, being spread out fairly evenly across all angles.

The consequences for the magnetic field in a slice through the centre of the simulation volume are shown in the two panels of Figure \ref{fig:m1200_bubble}. This figure shows a volume density slice through the centre of the simulation volume at 250 kyr and 325 kyr. A (1 pc)$^2$ window is centered on the star cluster in the left panel, and this window position is kept when plotting the second panel. Volume densities are coloured from a minimum at $n = 100$ cm$^{-3}$ to $n = 10^{6}$ cm$^{-3}$. We then overplot a magnetic vector map, using the magnetic field orientations in the plane of the slice, rather than performing a density-weighted projection as was done to generate Figures \ref{fig:m1200_large_scale_orientation} and \ref{fig:m500_large_scale_orientation}.

As the \HII region grows, the expanding bubble sweeps up a shell of material. The panel on the right of Figure \ref{fig:m1200_bubble} shows the magnetic field lines being swept up with this wall of material, consistent with other observations and theoretical work \citep[see, e.g.,][]{Lyutikov2006,DursiPfrommer2008,Arthur+2011,vanMarle+2015}. The magnetic field becomes compressed along the bubble wall, and its strength is enhanced by a factor of about 5--6, from around 20 Gauss to around 100--120 Gauss when comparing the B-field magnitude inside the shell versus just beyond outside it. Meanwhile, the magnetic field inside the bubbble is chaotic and disordered. We recall that in our 3D filaments analysis, Figure \ref{fig:m1200_rel_orientation_hist} showed how any coherent orientation of the magnetic field relative to the filamentary structure is largely erased by the end of the simulation (325 kyr), with the relative orientation approaching the random distribution. This is especially true for the highest-density gas, which is also where the massive stars formed and their radiative feedback injected the most kinetic energy.

\citet{Arthur+2011} also found that when an \HII region expands into a turbulent medium, the magnetic field tends to become ordered, lying parallel to the ionization front. This is consistent with what we observe in Figure \ref{fig:m1200_bubble}. They also reported that the magnetic field within the ionized region tended to be oriented perpendicular to the front, whereas in our case the field nearest the star cluster has the appearance of random orientation.

\begin{figure}
\includegraphics[width=88mm]{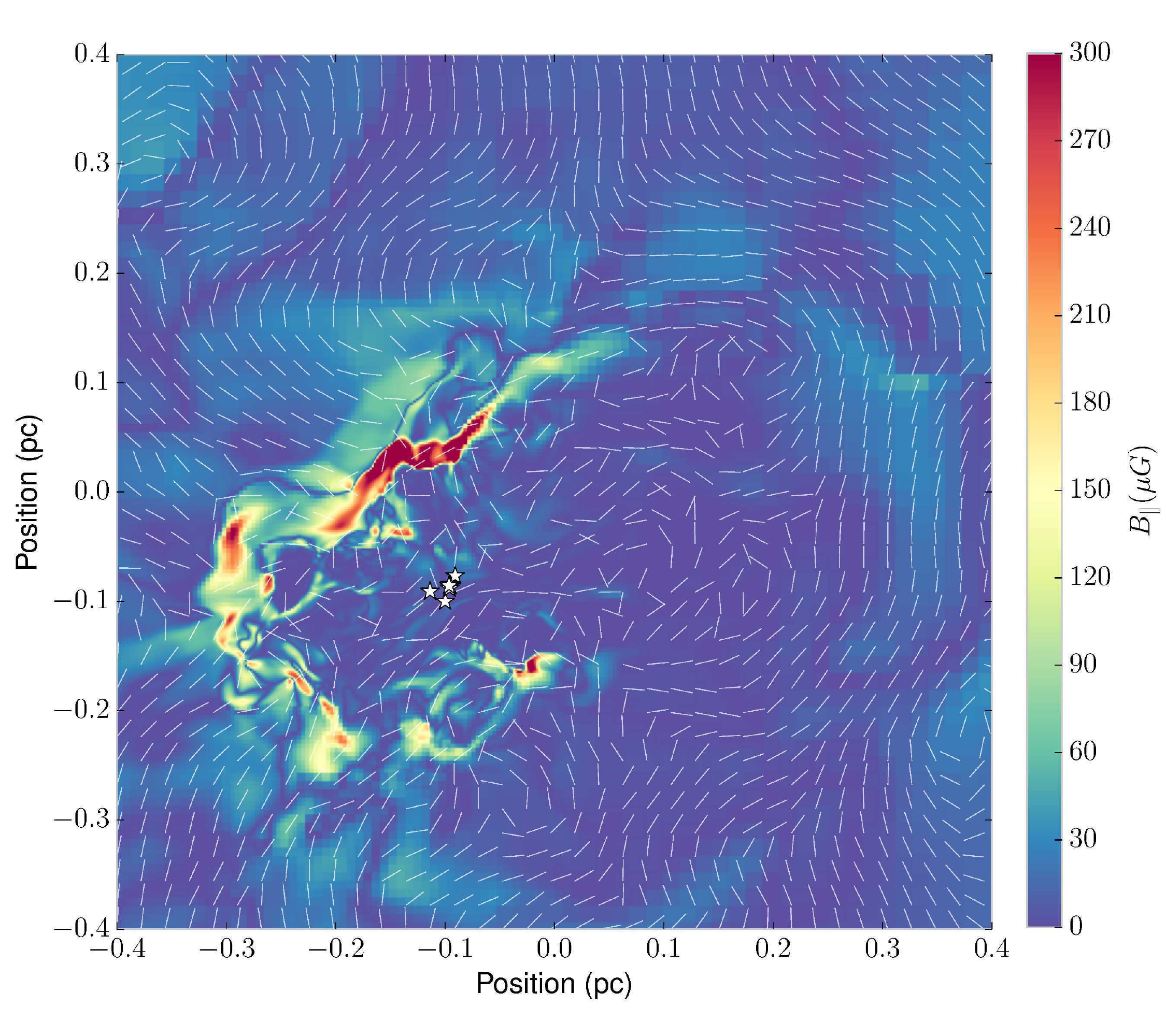}
\caption{A slice through the simulation volume of the {\tt MHD1200} simulation at 325 kyr showing the line-of-sight magnetic field strength around the \HII region driven by the cluster of stars in the lower-left corner of the image. Magnetic field vectors are overplotted giving the plane-of-sky magnetic field orientation.}
\label{fig:m1200_mag_los}
\end{figure}

The Rosette Nebula within the Monoceros molecular cloud is an example of an observed region with ongoing massive star formation that provided an opportunity to study the effect of an expanding \HII region within a magnetised environment. \citet{PlanckXXXIV} were able to fit an analytical model to this \HII region using \textit{Planck} 353 GHz dust polarisation data that was able to reproduce the observed rotation measure data. An enhancement of the line-of-sight magnetic field by about a factor of 4 was seen inside shell swept up by the \HII region.

Results from the \textit{Planck} mission offer opportunities to compare numerical simulations and high-resolution observations. We plotted the line-of-sight magnetic field strength from our {\tt MHD1200} simulation at 325 kyr, when the \HII region is already evolved. We show this in Figure \ref{fig:m1200_mag_los}. The plot centres on an area similar to that shown in the right panel of Figure \ref{fig:m1200_bubble}. We again overplot the plane-of-sky magnetic field vectors. The enhancement of the line-of-sight magnetic field strength in the shell of the expanding \HII region is roughly a factor of 4--10, depending on which part of the shell is probed. The magnetic field strength inside the \HII region varies over 1--20 $\mu G$, similar to the ambient field strength in our simulation and in agreement with the ambient field estimates in \citet{PlanckXXXIV}, though the studied \HII region in Rosette is much larger ($R \sim 20$ pc) than our simulated volume ($L \sim 4$ pc).

Massive stars are a possible mechanism for driving turbulence in molecular clouds. They also complicate the picture of how magnetic fields ought to orientated around filamentary clouds, randomizing it in some places, while possibly sweeping together field lines within expanding shells of material. We can expect sites of active star formation, especially massive star formation, to disturb the order of the magnetic field inherited from the ICM.

\section{Discussion}\label{sec:discussion}

In \citet{Kirk+2015}, we analyzed the properties of filaments resulting from hydrodynamic and magnetohydrodynamic simulations, finding that magnetic fields resulted in ``puffier'' filaments, i.e.~lower central densities, broader scale widths, and filaments less prone to gravitational fragmentation. The filamentary structure was extracted from column density projections, but no analysis was done on the relative orientation of the magnetic field and the filament. This was taken up by other authors \citep{SeifriedWalch2015}, who analyzed linear initial filament configurations and showed that perpendicular magnetic fields can result in filaments thinner than the proposed univeral filament width of 0.1 pc \citep{Arzoumanian+2011}. Different field configurations and turbulence will result in different fragmentation patterns.

One of the key differences between our simulations and those of \citet{SeifriedWalch2015} is the origin of the filaments. \citet{SeifriedWalch2015} starts with a single filament as an initial condition, whereas we form a network of filaments in a molecular cloud clump. In our simulations, filaments are the result of supersonic turbulence. In \citet{SeifriedWalch2015}, turbulence is present, but affects the fragmentation pattern of the gas; it is not responsible for the filament's structure.

\subsection{Magnetic fields, filament formation, and dynamics}

What is the case in nature? Do flows of gas along the magnetic fields of the intercloud medium result in filament-shaped clouds, or does supersonic turbulence help define both the filamentary structure and the magnetic field structure? Various scenarios for filament formation have been studied. In the supersonic turbulence scenario, colliding shocks create a network of filaments where dense gas can stagnate \citep[e.g.][]{Hartmann+2001,Padoan+2001,Arzoumanian+2011}. These colliding shocks may be driven by stellar feedback, supernovae, or other sources of turbulence. Our own use of a random, decaying initial supersonic velocity field with a turbulent power spectrum is motivated by this scenario. The use of driven turbulence would have continuously injected energy into the simulation, and would have been inappropriate for this type of study. The relative orientation of magnetic fields in our scenario will depend on the relative strengths of gravity, turbulence, and the magnetic field, as we have shown in this paper. For trans-Alfv\'{e}nic turbulence, gas compression can happen both perpendicular or parallel to magnetic fields.

A related scenario is colliding flows or cloud-cloud collisions \citep[see, e.g.][]{RedfieldLinsky2008,Banerjee2009,InoueFukui2013}. In the local ISM, cloud-cloud collisions may be responsible for the observed filamentary morphology \citep{RedfieldLinsky2008}. In cloud-cloud collisions, the magnetic fields may thread the massive, filamentary cloud cores perpendicular to the filaments \citep{InoueFukui2013}. Compression of Warm Neutral Medium (WNM) flows (possibly through turbulence) can trigger the condensation of cold gas structures, even filaments oriented parallel to magnetic fields as the shear of the turbulent flow stretches gas condensations into sheets and filaments \citep{AuditHennebelle2005,InoueInutsuka2009,Heitsch+2009,Saury+2014}. The ubiquity of filaments may thus be explained as generic turbulence sheers gas condensations into filaments, and magnetic fields may help keep these as coherent structures \citep{Hennebelle2013}. The orientation reported in \citet{PlanckXXXII} between matter structures in the diffuse ISM and magnetic fields could be a signature of the Cold Neutral Medium (CNM) filaments through turbulence.

Another mechanism for forming filamentary molecular clouds is B-field channeled gravitational contraction \citep[e.g.][]{Nakamura+Li2008}. Here Lorentz forces imply that gas motion along magnetic fields is unhindered, whereas gas moving perpendicular to field lines encouters a magnetic pressure. This means that gravity can channel gas along field lines, fragmenting the cloud into filaments that perendicular to the B-field, but parallel to each other.

In a related scenario, anisotropic sub-Alfv\'{e}nic turbulence has the effect of spreading gas preferentially along magnetic fields. In this case, filaments appear parallel to magnetic field lines. These latter two scenarios are considered for Gould Belt clouds by \citep{Li+2013}.

Which of these scenarios is true likely depends on the local environment: the relative strengths of turbulence and magnetic fields, the physical scales under consideration, the enclosed mass and boundedness of the region, the isotropy of the turbulence, and the star formation history of the region.

An important measure of the dynamical importance of a filament segment is the mass-per-unit-length, sometimes called the dynamical mass or line mass of a filament. An equilibrium analysis can be used to define a critical value for stability, as was done in \citet{Ostriker1964} who showed that for an isothermal cylinder, the mass per unit length is
\begin{equation}
m_{\mathrm{crit}} = \frac{2 c_s^2}{G} = \frac{2 k_B T}{\mu m_H G},
\end{equation}
where $c_s$ is the sound speed, $G$ is Newton's constant, $k_B$ is Boltzmann's constant, $T$ is the temperature, $\mu$ is the mean molecular weight, and $m_H$ is the mass of the hydrogen atom \citep[see also][]{Inutsuka+1997,FiegePudritz2000}. More generally, the total velocity dispersion should be used, since filaments are the products of supersonic motions and therefore may have nonthermal support, i.e.~$m_{\mathrm{crit}} = 2 \sigma^2 / G$ \citep{FiegePudritz2000}.

Line masses in excess of this critical value will undergo gravitational collapse and fragmentation to form protostars, with the most massive stars likely to be formed at the intersection of filaments \citep{Schneider+2012,Peretto+2013}. When using column density maps with traced filaments, one way of estimating the local mass-per-unit-length along filaments is to multiply the column density value by the characteristic filament width. \citet{Arzoumanian+2011} characterized the filaments in {\it Herschel} observations of IC 5146 as having a median width of $0.10 \pm 0.03$ pc. By estimating filament line masses in the column density projections from our simulation data, and pairing it with filament maps and magnetic field information, we can investigate whether relative orientation might be a function of the underlying line masses.  

\subsection{Effects of massive star formation}

In \citet{Klassen+2016}, we simulated the evolution of an isolated, massive protostellar core using a new hybrid radiative transfer code introduced in \citet{Klassen+2014}. In a core with a radius of 0.1 pc and an initial mass of $100 \Msun$, we were able to form a star of 16 $\Msun$ mass in around 30 kyr, which is much faster than what we see in this paper's simulations. This star then proceeded to grow to almost 30 $\Msun$ in another 10 kyr. The accretion rates in this paper are also an order of magnitude lower than in the isolated protostellar core simulation. What limits the accretion onto massive stars in the filamentary molecular cloud clump scenario? For one, turbulence may slow accretion by reconfiguring the gas reservoir into a network of filaments. Stars form along supercritical filaments, and initially have only these from which to draw mass. Accretion flows onto and along these filaments must then supply new material in order for the stars to continue growing. Magnetic fields, depending on their configuration, limit or enable these accretion flows, and provide additional support against gravitational collapse.

We use sink particles to represent stars as a practical necessity \citep{Krumholz+2004}. The size of the sink particle is ultimately set by the grid---it needs to have a radius of at least 2 grid cells in order to resolve the Jeans length with at least 4 cells \citep[the Truelove criterion,][]{Truelove+1997}. The sink particle radius in our {\tt MHD1200} simulation was set to $R_{\mathrm{sink}} = 1.758 \times 10^{16}$ cm, which is 1175 AU, or 3 grid cells. At this radius, the sink particle completely encloses the protostellar disk, through which much of the accretion takes place \citep{Kuiper+2011}. \citet{Beuther+2009} concluded from a study of 12 protostellar disk candidates, that the disks were fed from infalling outer envelopes and their radii were less than 1000 AU.

Ultimately, the mechanism that shut off accretion in the {\tt MHD1200} simulation was photoionization feedback. We did not include ionizing radiation in \citet{Klassen+2016}, which studied protostellar core collapse and disk accretion. Here, photoionization is cutting off the gas supplied by filamentary flow, strongly limiting the gas reservoir for the star cluster.

\section{Conclusions}\label{sec:conclusion}

We performed simulations of the evolution of turbulent, magnetised molecular clouds of various mean densities and \alfven Mach numbers close to unity. To each we applied the same turbulent velocity field, with an RMS Mach number of 6 in both cases. The mass-to-flux ratio was between 2 and 3, consistent with observations. We measured the virial parameter of each simulated cloud. The {\tt MHD500} simulation was initialized in virial balance, whereas the {\tt MHD1200} simulation had an initial virial parameter of 0.56, meaning that the gravitational binding energy substantially exceeded the kinetic energy.

The largest filaments formed in our simulation were on the order of 1--2 pc in size, i.e.~the size of our simulated region, and in each simulation we identified the primary structure, which we refer to as the main ``trunk'' filament using autocorrelation maps of the column density projection. We then measured the distribution of magnetic field vectors relative to the orientation of this primary filament.

We then applied the \disperse algorithm to extract the 3D filamentary structure from these simulations.  We trace along the filaments and measure the orientation of the filament and local physical variables, such as density and the magnetic field.

In summary, we make two major conclusions:
\begin{enumerate}
\item {\it The gravitational binding of a cloud has a profound effect on relative orientation of B-fields and dense filaments. For strongly bound clouds, we see the magnetic fields parallel to filaments in accretion flows along filaments.} For trans-Alfv\'{e}nic molecular clouds, coherent magnetic field structure depends on coherent velocity field structure. \textit{The filaments within them are largely the result of supersonic turbulence, not of slow accretion flows along magnetic field lines.} Most clouds are observed to have \alfven Mach numbers near unity. Simulations tend to focus on cases where the clouds are clearly sub-Alfv\'{e}nic or super-Alfv\'{e}nic, but we must also examine transition cases. There is no reason to expect magnetic fields to have large-scale coherent structure in these cases, but in clouds undergoing strong gravitational collapse, as in our {\tt MHD1200} simulation, which has a virial parameter of $\alpha_{\mathrm{vir}} = 0.56$, accretion flows onto the main filament result in a bimodal distribution of magnetic field orientation. Within the main filament, fields are aligned parallel to the long axis. Outside the main filament, magnetic fields partly show a perpendicular orientation relative to the main filament. The velocity field shows strong accretion flows perpendicular to and onto the main filament.

We compared the {\tt MHD1200} and {\tt MHD500} simulations at the same number of freefall times. The {\tt MHD1200} showed a preference for parallel orientation of the magnetic field relative to the main trunk filament, with accretion flows radially onto the filament and then along the filament axis towards the location of the star cluster. The {\tt MHD500}, which had a higher average mass density, showed a much more chaotic magnetic field, but with a trend towards a more perpendicular orientation.

\item {\it Radiation feedback from massive star formation disrupts the structure of both filaments and magnetic fields}. We looked at the effect of star formation and stellar radiative feedback in the {\tt MHD1200} simulation. Here we form a cluster of 7 massive stars, the most massive of which is about 16 $\Msun$ and has a luminosity of almost 23,000 $\Lsun$. The other stars in the cluster have masses of about $3 \Msun$ or below. The massive star dominates the others and drives the formation of an \HII region that appears as a blister on the side of the main trunk filament and expands outwards. This expanding bubble sweeps up a shell of gas and compresses the magnetic field, leading to an enhancement by a factor of 5--6. The magnetic field lines are seen to roughly trace the outline of the expanding shell. Within the bubble and in some parts outside the shell, the magnetic fields are chaotically orientated. Ultimately, the \HII region destroys the main trunk filament, cutting off the accretion flow onto the massive star. The relationship between the cumulative luminosity of the star cluster and the degree of cloud disruption would be an interesting area of future study.
\end{enumerate}

Additionally, we find that:
\begin{itemize}
\item {\it Highly bound clouds have a less random ordering of their magnetic fields than weakly bound clouds}. Our {\tt MHD500} simulation was more sub-Alfv\'{e}nic ($\mathcal{M}_A = 0.92$) than our {\tt MHD1200} simulation ($\mathcal{M}_A = 0.99$), yet it had the more disordered magnetic field structure. We attribute this to the cloud being in virial balance ($\alpha_{\mathrm{vir}} = 0.95$) as opposed to the very bound case of $\alpha_{\mathrm{vir}} = 0.56$ for the {\tt MHD1200} simulation. The kinetic energy of the cloud (including both thermal and non-thermal motions) was on par with both the magnetic field energy and the gravitational binding energy.

\item {\it At small-scale sub-parsec length filaments, the relative magnetic field structure is very complex}. The filamentary and magnetic field structure are influenced by the supersonic, turbulent velocity field, and the globally rotating molecular cloud clump also drags magnetic field lines into the plane of rotation. Over the course of the simulation, and within much less than a freefall time, the distribution of magnetic field orientations spreads out from an initially uniform field parallel with the $z$-axis to a broad range of angular values. 

The {\tt MHD500} simulation was more compact and had a higher average mass density. There did not appear a strong preference for orienting its filaments either parallel or perpendicular to the magnetic field as the simulation evolved. At late times and at lower density, some of the filaments did appear oriented perpendicular to the magnetic field, although this was not a strong trend.

\end{itemize}

In this trans-Alfv\'{e}nic regime, where magnetic energy balances turbulent kinetic energy, gravity's contribution to the energy budget is a determining factor in understanding how material is channeled onto filaments and the geometry of the magnetic field. We have studied an under-represented part of the parameter space and highlighted the importance of the virial parameter to be considered in tandem with the \alfven Mach number. Filament-aligned flow helps feed star clusters forming in dense regions within massive filaments, and their radiative feedback, especially via photoionization, may set the lifetimes of molecular cloud clumps. Magnetic fields certainly act to channel diffuse gas onto the main filament trunk and must finally be overcome by gravity if filamentary flow onto a forming cluster is to occur. 

\section*{Acknowledgments}

We thank an anonymous referee for a very useful report that helped to improve our manuscript. We also thank Henrik Beuther, Jouni Kainulainen, Thomas Henning, and Ralf Klessen for stimulating discussions. We would also like to thank Thierry Sousbie for having shared a pre-release version of the \disperse code with us. M.K.~acknowledges financial support from the Natural Sciences and Engineering Research Council (NSERC) of Canada. R.E.P.~is supported by an NSERC Discovery Grant. H.K.~was supported by a Banting Fellowship during the early stages of this project. R.E.P. thanks the MPIA and the Institut f\"{u}r Theoretische Astrophysik (ITA) of the Zentrum f\"{u}r Astronomie Heidelberg for support of his sabbatical leave (2015/16) when this project was completed.  

The \FLASH code was in part developed by the DOE-supported Alliances Center for Astrophysical Thermonuclear Flashes (ASCI) at the University of Chicago. This work was made possible by the facilities of the Shared Hierarchical Academic Research Computing Network (SHARCNET: www.sharcnet.ca) and Compute/Calcul Canada.

Much of the analysis and data visualization was performed using the {\tt yt} toolkit\footnote{http://yt-project.org} by \citet{ytpaper}.

\bibliography{magnetic_fields_orientation}

\label{lastpage}

\end{document}

%% file: table_sim_params.tex
\begin{table*}
\centering
\caption{Simulation parameters}
\begin{tabular*}{0.9\textwidth}{@{\extracolsep{\fill} } llccc}
\hline\noalign{\smallskip}
\multicolumn{5}{c}{Physical simulation parameters} \\
Parameter                   &             &                         & {\tt MHD500} & {\tt MHD1200}  \\
\noalign{\smallskip}\hline\noalign{\smallskip}
cloud radius                & [pc]        & $R_0$                   & 1.00        & 1.94        \\
total cloud mass            & [$\Msun$]     & $M_{\textrm{tot}}$      & 502.6       & 1209.2      \\
mean mass density           & [g/cm$^3$]  & $\bar{\rho}$            & 4.26e-21    & 1.39e-21    \\
mean number density         & [cm$^{-3}$] & $\bar{n}$               & 1188.98     & 388.841     \\
mean initial column density & [g/cm$^{-2}$] & $\bar{\Sigma}$        & 0.0262      & 0.0167      \\
mean initial column (number) density & [cm$^{-2}$] &                & $7.33 \times 10^{21}$ & $4.66 \times 10^{21}$ \\
mean molecular weight       &             & $\mu$                   & 2.14        & 2.14        \\
initial temperature         & [K]         & $T$                     & 10.0        & 10.0        \\
sound speed                 & [km/s]      & $c_{\textrm{s}}$        & 0.19        & 0.19        \\
1D velocity dispersion      & [km/s]      & $\sigma_{\textrm{1D}}$  & 0.64        & 0.55        \\
rms Mach number             &             & $\mathcal{M}_{\textrm{RMS}}$ & 6.0    & 6.0         \\
mean Mach number            &             & $\mathcal{M}$           & 4.36        & 3.71        \\
mean freefall time          & [Myr]       & $t_{\textrm{ff}}$       & 1.02        & 1.78        \\
sound crossing time         & [Myr]       & $t_{\textrm{sc}}$       & 10.4        & 20.2        \\
turbulent crossing time     & [Myr]       & $t_{\textrm{tc}}$       & 1.73        & 3.36        \\
magnetic field flux density & [$\mu$G]    & $\Phi$                  & 28.5        & 15.0        \\
Alfv\'{e}n speed            & [km/s]      & $v_A$                   & 1.23        & 1.13        \\
Alfv\'{e}n Mach number      &             & $\mathcal{M}_A$         & 0.92        & 0.99        \\
mass-to-flux ratio          &             & $\lambda$               & 2.33        & 2.82        \\
virial parameter            &             & $\alpha_{\textrm{vir}}$ & 0.95        & 0.56        \\ 
angular rotation frequency  & [s$^{-1}$]  & $\Omega_{\textrm{rot}}$ & $1.114 \times 10^{-14}$   & $1.114 \times 10^{-14}$   \\
rotational energy fraction  &             & $\beta_{\textrm{rot}}$  & 1.0 \%      & 3.2 \%      \\
critical thermal line mass  & [$\Msun$/pc]& $M_{\textrm{line}}^{\textrm{crit}}$ & 30.1   & 35.4      \\
critical turbulent line mass& [$\Msun$/pc]& $M_{\textrm{line}}^{\textrm{crit}}$  & 190.5  & 138.5      \\
\noalign{\smallskip}\hline\noalign{\smallskip}
\multicolumn{5}{c}{Numerical simulation parameters} \\
\noalign{\smallskip}\hline\noalign{\smallskip}
simulation box size         & [pc]        & $L_{\textrm{box}}$      & 2.0         & 3.89        \\
smallest cell size          & [AU]        & $\Delta x$              & 50.35       & 391.68      \\
\noalign{\smallskip}\hline\noalign{\smallskip}
\multicolumn{5}{c}{Simulation outcomes} \\
\noalign{\smallskip}\hline\noalign{\smallskip}
final simulation time       & [kyr]      & $t_{\textrm{final}}$ & 161.3 kyr & 326.4 \\[1mm] 
\hline \\
\end{tabular*}
\label{table:simulation_parameters}
\end{table*}